\documentclass[zpreprint,zbstdefault]{zeus_paper}
\usepackage[english]{babel}

\newcommand{\ZcoosysB}{%
The ZEUS coordinate system is a right-handed Cartesian system, with the $Z$
axis pointing in the proton beam direction, referred to as the ``forward
direction'', and the $X$ axis pointing left towards the centre of HERA.
The coordinate origin is at the nominal interaction point.\xspace}
\newcommand{\Zpsrap}{%
The pseudorapidity is defined as $\eta=-\ln\left(\tan\frac{\theta}{2}\right)$,
where the polar angle, $\theta$, is measured with respect to the proton beam
direction.\xspace}

\newcommand{\ZcoosysfnBeta}{\footnote{\ZcoosysB\Zpsrap}}
\newcommand{\Zdetdesc}{%
A detailed description of the ZEUS detector can be found 
elsewhere~\cite{zeus:1993:bluebook}. A brief outline of the 
components 
most relevant for this analysis is given
below.\xspace}
\newcommand{\Zctddesc}[1]{%
Charged particles were tracked in the central tracking detector (CTD)~\citeCTD,
which operated in a magnetic field of $1.43\Tesla$ provided by a thin 
superconducting coil. The CTD consisted of 72~cylindrical drift chamber 
layers, organised in 9~superlayers covering the polar-angle#1 region 
\mbox{$15^\circ<\theta<164^\circ$}. The transverse-momentum resolution for
full-length tracks was $\sigma(p_T)/p_T=0.0058p_T\oplus0.0065\oplus0.0014/p_T$,
with $p_T$ in $\Gev$.}
\newcommand{\Zcaldesc}{%
The high-resolution uranium-scintillator calorimeter (CAL)~\citeCAL consisted 
of three parts: the forward (FCAL), the barrel (BCAL) and the rear (RCAL)
calorimeters. Each part was subdivided transversely into towers and
longitudinally into one electromagnetic section (EMC) and either one (in RCAL)
or two (in BCAL and FCAL) hadronic sections (HAC). The smallest subdivision of
the calorimeter is called a cell.  The CAL energy resolutions, as measured under
test-beam conditions, were $\sigma(E)/E=0.18/\sqrt{E}$ for electrons and
$\sigma(E)/E=0.35/\sqrt{E}$ for hadrons, with $E$ in $\Gev$.}

\chardef\usc=95
\chardef\til=126
\catcode`\@=11 
\DeclareRobustCommand\xdotspace{\futurelet\@let@token\@xdotspace}
\def\@xdotspace{%
  \ifx\@let@token.\else
  \ifx\@let@token\bgroup.\else
  \ifx\@let@token\egroup.\else
  \ifx\@let@token\/.\else
  \ifx\@let@token\ .\else
  \ifx\@let@token~.\else
  \ifx\@let@token!.\else
  \ifx\@let@token,.\else
  \ifx\@let@token:.\else
  \ifx\@let@token;.\else
  \ifx\@let@token?.\else
  \ifx\@let@token/.\else
  \ifx\@let@token'.\else
  \ifx\@let@token).\else
  \ifx\@let@token-.\else
  \ifx\@let@token\@xobeysp.\else
  \ifx\@let@token\space.\else
  \ifx\@let@token\@sptoken.\else
   .\space
   \fi\fi\fi\fi\fi\fi\fi\fi\fi\fi\fi\fi\fi\fi\fi\fi\fi\fi}
\catcode`\@=12 

\newcommand{\stru}[2]{%
   \relax\ifmmode\hbox{\vrule height#1 depth#2 width0pt}%
   \else\vrule height#1 depth#2 width0pt\fi}

\newcommand{\Ronum}[1]{\uppercase\expandafter{\romannumeral#1}}
\newcommand{\ronum}[1]{\expandafter{\romannumeral#1}}
\DeclareRobustCommand{\LaTeXZ}{%
  \LaTeX\kern-.05em4\kern-.1em
  {\raisebox{-0.2ex}{$\scriptstyle\text{ZEUS}$}}\xspace}

\DeclareMathAlphabet{\mathbf}{OT1}{cmr}{bx}{sl}
\newcommand{\eVdist}{\kern-0.06667em}

\newcommand{\Gev}{{\text{Ge}\eVdist\text{V\/}}}

\newcommand{\gev}{{\,\text{Ge}\eVdist\text{V\/}}}

\newcommand{\Tesla}{\,\text{T}}

\newcommand{\slashfrac}[2]{%
  \raisebox{0.5ex}{\ensuremath #1}\kern-0.12em/\kern-0.08em
  \raisebox{-.8ex}{\ensuremath #2}}

\newcommand{\sqr}[3]{%
    {\vcenter{\hrule height.#3ex\hbox{\vrule width.#2ex height#1ex
     \kern#1ex\vrule width.#3ex}\hrule height.#2ex}}}

\newcommand{\widebar}[1]{%
   \mkern1.5mu\overline{\mkern-1.5mu#1\mkern-1.mu}\mkern1.mu}
\catcode`\@=11 
\newcommand{\parenbar}{\mathpalette\p@renb@r}
\def\p@renb@r#1#2{\vbox{%
  \ifx#1\scriptscriptstyle \dimen@.7em\dimen@ii.2em\else
  \ifx#1\scriptstyle \dimen@.8em\dimen@ii.25em\else
  \dimen@1em\dimen@ii.4em\fi\fi \offinterlineskip
  \ialign{\hfill##\hfill\cr
    \vbox{\hrule width\dimen@ii}\cr
    \noalign{\vskip-.3ex}%
    \hbox to\dimen@{$\mathchar300\hfil\mathchar301$}\cr
    \noalign{\vskip-.3ex}%
    $#1#2$\cr}}}
\catcode`\@=12 

\newcommand{\pbar}{\widebar{p}}

\newcommand{\IP}{{\rm I$\kern-0.01667em$P}\xspace}

\mathchardef\qsm=63
\mathchardef\pls=43
\mathchardef\mns=512
\mathchardef\plm=518
\mathchardef\eql=61
\mathchardef\smallleft=300
\mathchardef\smallright=301
\mathchardef\les=316
\mathchardef\gre=318
\mathchardef\leq=532
\mathchardef\grq=533

\catcode`\@=11 
\newcounter{pict@width}
\newcounter{pict@height}
\newlength{\pict@scale}
\newcommand{\psfigadd}[4]{%
\setcounter{pict@width}{1*\ratio{#2+\pict@scale/2}{\pict@scale}}
\setcounter{pict@height}{1*\ratio{#3+\pict@scale/2}{\pict@scale}}
\setlength{\unitlength}{\pict@scale}
\hbox to #2{\hspace{-\fill}\begin{picture}(\thepict@width,\thepict@height)
\put(0,0){\psfig{figure=#1,width=#2,height=#3,clip=}}
\SetScale{0.283466457}
\SetWidth{1.763889}
{#4}
\end{picture}}
}
\newcounter{pict@widthfst}
\newcounter{pict@widthscd}
\newcounter{pict@widthtot}
\newcommand{\psfigaddtwo}[7]{%
\setcounter{pict@widthfst}{1*\ratio{#2+\pict@scale/2}{\pict@scale}}
\setcounter{pict@widthscd}{1*\ratio{#2+#4+\pict@scale/2}{\pict@scale}}
\setcounter{pict@widthtot}{1*\ratio{#2+#4+#6+\pict@scale/2}{\pict@scale}}
\setcounter{pict@height}{1*\ratio{#3+\pict@scale/2}{\pict@scale}}
\setlength{\unitlength}{\pict@scale}
\hbox{\hspace{-\fill}\begin{picture}(\thepict@widthtot,\thepict@height)
\put(0,0){\psfig{figure=#1,width=#2,height=#3,clip=}}
\put(\thepict@widthscd,0){\psfig{figure=#5,width=#6,height=#3,clip=}}
\SetScale{0.283466457}
\SetWidth{1.763889}
{#7}
\end{picture}}
}
\newcommand{\psfigror}[4]{%
\setcounter{pict@width}{1*\ratio{#2+\pict@scale/2}{\pict@scale}}
\setcounter{pict@height}{1*\ratio{#3+\pict@scale/2}{\pict@scale}}
\setlength{\unitlength}{\pict@scale}
\hbox{\begin{picture}(\thepict@width,\thepict@height)
\put(0,\thepict@height){\psfig{figure=#1,width=#3,height=#2,clip=,angle=270}}
\SetScale{0.283466457}
\SetWidth{1.763889}
{#4}
\end{picture}}
}
\newcommand{\psfigrol}[4]{%
\setcounter{pict@width}{1*\ratio{#2+\pict@scale/2}{\pict@scale}}
\setcounter{pict@height}{1*\ratio{#3+\pict@scale/2}{\pict@scale}}
\setlength{\unitlength}{\pict@scale}
\hbox{\begin{picture}(\thepict@width,\thepict@height)
\put(0,0){\psfig{figure=#1,width=#3,height=#2,clip=,angle=90}}
\SetScale{0.283466457}
\SetWidth{1.763889}
{#4}
\end{picture}}
}
\catcode`\@=12 
\newlength\listtextwidth

\catcode`\@=11 
\newlength{\@tabfninsert}
\newlength{\@tabfnwidth}
\newcommand{\tabfootnote}[2]{%
  \setlength{\@tabfninsert}{0.8em}
  \setlength{\@tabfnwidth}{\textwidth}
  \addtolength{\@tabfnwidth}{-\@tabfninsert}
  \addtolength{\@tabfnwidth}{-0.4em}
  \noindent\makebox[\@tabfninsert][r]{\footnotesize$^{#1}$\hfil}\hfill%
  \parbox[t]{\@tabfnwidth}{\footnotesize #2\hfill}}
\catcode`\@=12 

\newcommand{\pythia}{\textsc{Pythia}\xspace}
\newcommand{\rapgap}{\textsc{Rapgap}\xspace}

\newcommand{\bbbar}{\ensuremath{b\bar{b}}\xspace}

\newcommand{\mal}{\!\cdot\!}

\newcommand{\jpsi}{\ensuremath{J/\psi}\xspace}

\def\citeCTD{{\cite{%
nim:a279:290,*npps:b32:181,*nim:a338:254%
}}\xspace}
\def\citeCAL{{\cite{%
nim:a309:77,*nim:a309:101,*nim:a321:356,*nim:a336:23%
}}\xspace}

\begin{document}
\prepnum{DESY--08--129}

\title{
Measurement of beauty production from dimuon events at HERA
}                                                       
                    
\author{ZEUS Collaboration}
\date{23 September 2008}

\abstract{
Beauty production in events containing two muons in the final
state has been measured with the ZEUS detector at HERA using an integrated
luminosity of $114~$pb$^{-1}$.
A low transverse-momentum threshold for muon identification, in combination
with the large rapidity coverage of the ZEUS muon system, gives access to
almost the full phase space for beauty production.
The total cross section for beauty production in $ep$ collisions at 
$\sqrt{s}= 318$~GeV has been measured to be
$\sigma_{\rm tot}(ep \to b\bar b X) = 
   13.9 \pm 1.5({\rm stat.}) ^{+4.0}_{-4.3}({\rm syst.}) {\rm\ nb}. $
Differential cross sections and a measurement of $b\bar{b}$ correlations 
are also obtained, and compared
to other beauty cross-section measurements, Monte Carlo models and 
next-to-leading-order QCD
predictions. 
}

\makezeustitle

\def\3{\ss}                                                                                        
\pagenumbering{Roman}                                                                              
                                                   %
\begin{center}                                                                                     
{                      \Large  The ZEUS Collaboration              }                               
\end{center}                                                                                       
  S.~Chekanov,                                                                                     
  M.~Derrick,                                                                                      
  S.~Magill,                                                                                       
  B.~Musgrave,                                                                                     
  D.~Nicholass$^{   1}$,                                                                           
  \mbox{J.~Repond},                                                                                
  R.~Yoshida\\                                                                                     
 {\it Argonne National Laboratory, Argonne, Illinois 60439-4815, USA}~$^{n}$                       
\par \filbreak                                                                                     
  M.C.K.~Mattingly \\                                                                              
 {\it Andrews University, Berrien Springs, Michigan 49104-0380, USA}                               
\par \filbreak                                                                                     
  P.~Antonioli,                                                                                    
  G.~Bari,                                                                                         
  L.~Bellagamba,                                                                                   
  D.~Boscherini,                                                                                   
  A.~Bruni,                                                                                        
  G.~Bruni,                                                                                        
  F.~Cindolo,                                                                                      
  M.~Corradi,                                                                                      
\mbox{G.~Iacobucci},                                                                               
  A.~Margotti,                                                                                     
  R.~Nania,                                                                                        
  A.~Polini\\                                                                                      
  {\it INFN Bologna, Bologna, Italy}~$^{e}$
% ...
\newpage                                                                                           
\enlargethispage{5cm}                                                                              
$^{\    1}$ also affiliated with University College London,                                        
United Kingdom\\                                                                                   
% ...
$^{\    5}$ This work was supported in part by the Marie Curie Actions Transfer of Knowledge       
project COCOS (contract MTKD-CT-2004-517186)\\                                                     
% ...
%
\newpage   
                                                           %
                                                           %
\begin{tabular}[h]{rp{14cm}}                                                                       
$^{a}$ &  supported by the Natural Sciences and Engineering Research Council of Canada (NSERC) \\  
$^{b}$ &  supported by the German Federal Ministry for Education and Research (BMBF), under        
          contract numbers 05 HZ6PDA, 05 HZ6GUA, 05 HZ6VFA and 05 HZ4KHA\\                         
$^{c}$ &  supported in part by the MINERVA Gesellschaft f\"ur Forschung GmbH, the Israel Science   
          Foundation (grant no. 293/02-11.2) and the U.S.-Israel Binational Science Foundation \\  
$^{d}$ &  supported by the Israel Science Foundation\\                                             
$^{e}$ &  supported by the Italian National Institute for Nuclear Physics (INFN) \\                
$^{f}$ &  supported by the Japanese Ministry of Education, Culture, Sports, Science and Technology 
          (MEXT) and its grants for Scientific Research\\                                          
$^{g}$ &  supported by the Korean Ministry of Education and Korea Science and Engineering          
          Foundation\\                                                                             
$^{h}$ &  supported by the Netherlands Foundation for Research on Matter (FOM)\\                   
$^{i}$ &  supported by the Polish State Committee for Scientific Research, project no.             
          DESY/256/2006 - 154/DES/2006/03\\                                                        
$^{j}$ &  partially supported by the German Federal Ministry for Education and Research (BMBF)\\   
$^{k}$ &  supported by RF Presidential grant N 8122.2006.2 for the leading                         
          scientific schools and by the Russian Ministry of Education and Science through its      
          grant for Scientific Research on High Energy Physics\\                                   
$^{l}$ &  supported by the Spanish Ministry of Education and Science through funds provided by     
          CICYT\\                                                                                  
$^{m}$ &  supported by the Science and Technology Facilities Council, UK\\                         
$^{n}$ &  supported by the US Department of Energy\\                                               
$^{o}$ &  supported by the US National Science Foundation. Any opinion,                            
findings and conclusions or recommendations expressed in this material                             
are those of the authors and do not necessarily reflect the views of the                           
National Science Foundation.\\                                                                     
$^{p}$ &  supported by the Polish Ministry of Science and Higher Education                         
as a scientific project (2006-2008)\\                                                              
$^{q}$ &  supported by FNRS and its associated funds (IISN and FRIA) and by an Inter-University    
          Attraction Poles Programme subsidised by the Belgian Federal Science Policy Office\\     
$^{r}$ &  supported by the Malaysian Ministry of Science, Technology and                           
Innovation/Akademi Sains Malaysia grant SAGA 66-02-03-0048\\                                       
\end{tabular}                                                                                      
                                                           %
                                                           %

\pagenumbering{arabic} 
\pagestyle{plain}

\section{\label{sec-int}Introduction}

The production of beauty quarks in $ep$ collisions at HERA provides a
stringent test of perturbative Quantum Chromodynamics (QCD), since the
large $b$-quark mass ($m_b \sim 5 \gev$) gives a hard scale that
should ensure reliable predictions in all regions of phase space,
including the kinematic threshold. Especially
in this region, with $b$-quark transverse momenta comparable to or less 
than the $b$-quark mass,
next-to-leading-order (NLO) QCD calculations in which the (massive)
$b$ quarks are
generated dynamically are expected to provide accurate predictions
\cite{np:b412:225,*np:b454:3-24,*frixione3,
      np:b452:109,*pl:b353:535,*pr:d57:2806}.

The cross section for beauty production has been measured in $p\pbar$
collisions at the S$p\pbar$S
\cite{beautyUA10,*beautyUA1a,*beautyUA1,*beautyUA1b}
and Tevatron colliders
\cite{beautyCDF1,*beautyCDF2,*beautyCDF3,*beautyCDF4,*beautyCDF4a,*beautyCDF5,
*beautyCDF5a,*beautyCDF7,
*beautyCDF8,*beautyCDF9,*beautyD00,*beautyD01,*beautyD02,*beautyD03},
in $\gamma \gamma$ interactions at LEP 
\cite{beautyLEP1,*beautyLEP3,beautyLEP4}, and in
fixed-target $\pi N$ \cite{WA78,*E706} and $pN$
\cite{E771,*E789,*HERAB} experiments.
Most results, including recent results from the Tevatron, are in good
agreement with QCD predictions. Large discrepancies are observed in some 
\cite{beautyLEP1} of the results from $\gamma\gamma$ interactions at LEP.

In most of the previous measurements of beauty production at HERA,
beauty events were selected by requiring the presence of one or more
jets, tagged by a muon or electron from the semi-leptonic decay of one
of the $b$ quarks
\cite{pl:b467:156,*epj:c18:625,epj:c41:453,pr:d70:012008,pl:b599:173,desy-08-056}, 
or by tracks originating from the secondary decay vertex of beauty 
hadrons \cite{epj:c40:349,*epj:c45:23,*H1phjets}.
This restricts the measurements to $b$ quarks with high transverse
momentum ($p_T$).

This paper reports measurements of beauty production via the reaction
$ep\rightarrow \bbbar X\rightarrow \mu\mu X'$ using the ZEUS detector
at HERA. The dimuon final state yields a data sample enriched in 
$b \bar b$ pairs, and with strongly suppressed backgrounds from other
processes. This allows low muon-$p_T$ ($p_T^\mu$) thresholds to be applied 
without any jet requirements, and gives access
to a larger region of phase space, especially towards lower transverse momenta
of the $b$~quarks.

Conceptually, the analysis is similar to the H1 and ZEUS analyses of 
beauty in $D^*\mu$ final 
states \cite{pl:b621:56-71,epj:c50:1434}, with  three significant differences.
The larger branching ratio yields higher statistics, so that differential 
cross sections can be measured. The wider rapidity 
coverage and very low $p_T$ threshold allow the extraction of the total
beauty cross section with little extrapolation. The low charm background 
in the dimuon final state, partially due to the harder $b$ fragmentation,
allows measurements of $b\bar b$ correlations, testing the 
influence of higher-order contributions on the perturbative calculations. 

\section{\label{sec-exp}Experimental set-up}
The data sample used in this analysis corresponds to an integrated luminosity
${\cal L}=114.1 \pm 2.3 ~\rm{pb}^{-1}$, collected with the ZEUS detector
from 1996 to 2000.
In 1996--97,
HERA provided collisions between an electron\footnote{Electrons 
and positrons are both referred to as electrons in this paper.}
beam of  $E_e=27.5 \gev$ and a proton beam of
$E_p=820\gev$, corresponding to a  centre-of-mass energy  $\sqrt s=300\gev$
(${{\cal L}_{300}}=38.0\pm 0.6~ \rm{pb}^{-1}$). In
1998--2000, the proton-beam energy was
$E_p=920\gev$,  corresponding to $\sqrt s=318\gev$
(${{\cal L}_{318}}=76.1\pm 1.7~\rm{pb}^{-1}$).

\Zdetdesc

\Zctddesc\ZcoosysfnBeta

\Zcaldesc

The muon system consisted of rear, barrel (R/BMUON)~\cite{brmuon}
and forward (FMUON)~\cite{zeus:1993:bluebook} tracking detectors. \
The B/RMUON consisted of 
limited-streamer (LS) tube chambers placed behind the BCAL (RCAL), inside 
and outside the magnetised iron yoke surrounding the CAL. The barrel and 
rear muon chambers covered polar angles from 34$^{\circ}$ to 135$^{\circ}$ 
and from 135$^{\circ}$ to 171$^{\circ}$, respectively. 
The FMUON consisted of six planes of LS tubes and
four planes of drift chambers covering the angular region from 
5$^{\circ}$ to 32$^{\circ}$. 
The muon system exploited the magnetic field of the iron yoke and, in the 
forward direction, of two iron toroids magnetised to  1.6~T to provide an 
independent measurement of the muon momentum.

Muons were also detected by the sampling Backing Calorimeter (BAC) \cite{BACP}.
This detector consisted of 5200 
proportional drift chambers which were typically 5~m long and had a wire 
spacing of 1~cm. The chambers were inserted into the iron yoke of the ZEUS 
detector (barrel and two end caps) covering the CAL. The BAC was equipped with 
analogue readout for energy measurement and digital readout for muon tracking.
The former was based on 2000 towers ($50 \times 50$ cm$^2$), 
providing an energy
resolution of  $\sim 100\%/\sqrt E$. The digital information from the wires
allowed the reconstruction of muon trajectories in two dimensions ($XY$ in 
barrel and $YZ$ in end caps) with an accuracy of a few mm.

\section{\label{sec:meas_principle}Principle of the measurement}
Events with at least two muons in the final state were selected.
Two principal event classes contribute to the beauty 
signal to be measured. 
The first consists of events in which the two muons originate
from the same parent $b$ quark\footnote{
Unless stated otherwise, throughout this paper, the term $b$ quark 
includes $\bar b$.}, 
e.g. through the sequential decay chain 
$b \to c \mu X \to s \mu \mu X'$. 
These yield unlike-sign muon pairs produced in the same event hemisphere
and with dimuon invariant masses of 
$m^{\mu\mu}_{\rm inv} < 4~\gev$ 
(i.e. a partially reconstructed $B$-meson mass). 
The second class consists of events in which the two muons originate from
different beauty quarks of a $b \bar b$ pair. These can yield both like- and 
unlike-sign dimuon
combinations, depending on whether the muon originates from the decay
of the primary beauty quark, or from a secondary charm quark. 
In addition, $B^0 \bar B^0$ mixing can dilute these charge correlations.
Muons from different $b$ quarks will predominantly be 
produced in different hemispheres, and tend to have a large dimuon mass.

An important background contribution
arises from primary charm-quark pair production where both
charm quarks decay into a muon. This yields
unlike-sign muon pairs only, with the two muons
produced predominantly in opposite hemispheres. 
Since this background is too small to be measured directly from the 
dimuon data, it was normalised to the charm contribution as determined from
the ZEUS $D^*+\mu$ sample \cite{epj:c50:1434}
which has a similar event topology and covers a similar though 
somewhat more restricted kinematic range.

Other backgrounds yielding unlike-sign muon pairs include heavy quarkonium 
decays and Bethe-Heitler (BH) processes. In contrast to muons from 
semileptonic decays,
muons from these sources are not directly accompanied by hadronic activity,
thus giving an isolated muon signature.

Beauty production is the only source of genuine like-sign muon pairs. 
Background contributions to both like- and unlike-sign combinations 
include events in which either one or both muons are false, i.e. originate 
from $K \to \mu$ or $\pi \to \mu$ decays in flight or are misidentified 
hadrons. 
Studies \cite{thesis:bloch:2005-tmp-43fb5be7} have shown that
the charges of such false-muon pairs
are almost uncorrelated, i.e. the contributions
to the like- and unlike-sign dimuon distributions are almost equal.
The  difference between the unlike- ($N_{\rm data}^{\rm u}$) and 
like-sign ($N_{\rm data}^{\rm l}$) 
distributions is thus
essentially free from false-muon background, without the need to simulate this 
background using Monte Carlo (MC) methods. 
Once the background contributions from open charm ($N_{\rm charm}$), 
$J/\psi$ and 
other heavy vector mesons ($N_{\rm VM}$) and Bethe-Heitler ($N_{\rm BH}$) 
are known, this difference
can be used to measure the beauty 
contribution $N_{b\bar{b} \to \mu\mu}$ according to the formula
\begin{equation}
  \label{eq: n_beauty_calc}
N_{b\bar{b} \to \mu\mu} = \left( N_{\rm data}^{\rm u} -  N_{\rm data}^{\rm l}
                       - \left( N_{\rm charm} + N_{\rm VM} + N_{\rm BH} \right)
                         \right)
                         \times \left(
                         \frac{N_{b\bar{b}}^{\rm u} + N_{b\bar{b}}^{\rm l} }
                              {N_{b\bar{b}}^{\rm u} - N_{b\bar{b}}^{\rm l} }
                              \right)^{\rm MC} 
\end{equation}
where the last term refers to the unlike-sign ($N_{b\bar{b}}^{\rm u}$) and 
like-sign ($N_{b\bar{b}}^{\rm l}$) beauty contributions predicted by the MC.
Small corrections to this procedure will be explained in Section \ref{sect:sigex}.
The beauty signal is hence extracted from the 
difference between the unlike- and like-sign samples.

The like-sign false-muon background can then be obtained from the data by 
subtracting the MC like-sign beauty contribution, properly scaled to the 
measurement, from the total like-sign sample, while the unlike-sign background 
is a simple reflection of the like-sign background.
This method to obtain the false-muon background contributions 
will be referred to as the subtraction method.

Since one of the goals is the determination of the total beauty production
cross section in $ep$ collisions, events from deep inelastic scattering 
(DIS), where the photon virtuality, $Q^2$, is larger than 1\,GeV$^2$, and photoproduction
($Q^2 < 1$ GeV$^2$) were not explicitly separated.

The average cross sections obtained from the two different running periods 
($\sqrt{s}=$ 300 and 318 GeV) are all expressed in terms of a single cross 
section at $\sqrt{s}=$ 318 GeV. This involves a typical correction of +2\%.

\section{\label{sec-sel}Event selection and reconstruction}

\subsection{\label{sec-sel-trig}   Trigger selection}

The data were selected online by means of a three-level trigger system
\cite{zeus:1993:bluebook,uproc:chep:1992:222} through an
inclusive ``or'' of four different trigger channels:
\begin{itemize}
\item a muon reaching the inner B/RMUON chambers and matched to a minimum
      ionising energy deposit (MIP) in the CAL or any muon
      reaching the outer B/RMUON chambers (muon channel);
\item a reconstructed $D$ meson candidate ($D^*$ channel
\cite{np:b729:492-525}, plus similar chains for other charm mesons
\cite{epj:c44:13});
\item two jets (dijet channel \cite{pr:d70:012008}); 
\item a scattered-electron candidate in the CAL (DIS channel \cite{pl:b599:173}).
\end{itemize}
For part of the data taking, the requirements on the DIS and dijet channels 
were loosened in the presence of any muon in the inner B/RMUON chambers.
The non-muon triggers were used to gain geometric acceptance for 
regions not covered by the B/RMUON chambers, and to evaluate the efficiency 
of the muon triggers.
Owing to this redundancy, the trigger efficiency for dimuon events
with reconstructed muons from beauty was high, $80 \pm 4$\%.

\subsection{\label{sec-sel-cal}  Event selection}

The large mass of a $b\bar b$ pair, at least $\sim 10$ GeV, usually leads 
to a significant amount
of energy deposited in the more central parts of the detector.
To suppress backgrounds from false-muon events and charm, a hadronic 
transverse energy cut
$$ 
  E_T \geq 8~\gev 
$$ 
was applied, where
$$ 
  E_T = 
\begin{cases}
 E_T^{\theta>10^\circ}  & \mathrm{no~ scattered~ electron}\\
 E_T^{\theta>10^\circ} - E_T^e & \mathrm{with~ scattered~ electron}.
\end{cases}
$$ 
The transverse energy was calculated as 
$E_T^{\theta>10^\circ}= \Sigma_{i,\theta_i>10^\circ}(E_i \sin \theta_i)$, where
the sum runs over all energy deposits in the CAL with the polar angle above 
$10^\circ$. The latter restriction is imposed to remove proton-remnant effects.
If detected, 
the energy of the scattered electron ($E_T^e$) was subtracted. 
The detection criteria for the scattered electron were the same as in 
a previous publication \cite{epj:c50:1434}.

Various tracking requirements were 
imposed \cite{thesis:bloch:2005-tmp-43fb5be7}, the most important of which was
that the reconstructed longitudinal vertex position should be consistent with 
an $ep$ interaction, $|Z_{\rm vtx}| < 50$ cm.  

\subsection{\label{sec-sel-mu}   Muon selection}

Muons were reconstructed offline using an inclusive ``or'' of the 
following procedures:
\begin{itemize}
\item
a muon track was found in the inner B/RMUON chambers.
A match in position and angle to a CTD track was required.
In the bottom region, where no inner chambers are present, the outer chambers
were used instead. If a match was found to both inner and outer chambers,
a momentum-matching criterion was added;
\item
a muon track was found in the FMUON chambers. 
Within the CTD acceptance, a match in position and angle to a CTD track was 
required and the momentum was obtained from a combined fit of the 
CTD and FMUON information.
Outside the CTD acceptance, candidates well measured in FMUON only
and fitted to the primary vertex were accepted;
\item 
a muon track or localised energy deposit was found in the BAC, and
matched to a CTD track, from which the muon momentum was obtained. In the 
forward region of the detector, a MIP
in the calorimeter was required in addition in order to reduce background
related to the proton beam or to the punch-through of high energy hadrons.   
\end{itemize}
Most muons are within the geometric acceptance of more than one of these 
algorithms. 
The overall efficiency is about 80\%
for high-momentum muons (more than 2-5 GeV, depending on $\eta$).

Two different kinematic selections were made.
In the barrel region, the requirement that the muons reach at least the 
inner muon chambers implies a muon transverse momentum ($p_T^\mu$) of 
about 1.5 GeV 
or more. In order to have uniform kinematic acceptance, a cut 
$$ 
p_{T}^{\mu}>1.5\gev
$$ 
was therefore applied to all muons (selection A).

In the forward and rear regions, lower $p_T$ muons can be detected,
although with somewhat higher background. To cover the largest possible phase 
space for the intended measurement of a total beauty-production cross section, 
the $p_T$ cut was lowered to   
$$ 
p_{T}^{\mu}>0.75\gev
$$ 
for high-quality muons \cite{thesis:bloch:2005-tmp-43fb5be7}, 
i.e. muons seen by more than one muon detector and/or 
confirmed by a MIP in the CAL (selection B). 
For other muons satisfying all previously 
listed criteria, the cut $p_{T}^{\mu}>1.5\gev$ was retained to
keep the background low. 
Selection A is thus a subset of selection B.

At least two such muon candidates were required per event.
No explicit cut on the muon angle was applied for either selection. 
The angular coverage 
of the muon chambers, BAC and CTD gives continuous useable
acceptance in the pseudorapidity region
$$ 
-2.2\lesssim \eta^{\mu}\lesssim 2.5 \ .
$$ 

To suppress events with ambiguous matches between CTD tracks and muon 
chamber segments as well as genuine dimuons from prompt light-meson decays
(e.g. $\rho \to \mu\mu$), 
a dimuon invariant mass ($m^{\mu\mu}$) cut of 
$$ 
  m^{\mu\mu} >  1.5~\gev
$$ 
was applied.
This implies a minimum opening angle between the two muons. 

Events with a very forward and a very backward muon candidate, a topology 
not favoured for the beauty signal,
were removed by a cut on the difference in pseudorapidity of the 
two muon candidates of
$$ 
  |\eta^{\mu_1}-\eta^{\mu_2}|<3.0.
$$ 

Muon candidates with badly measured momentum 
(predominantly from false-muon backgrounds)
were suppressed using the 
imbalance between the transverse momenta of the muons
$$ 
(|p_T^{\mu_1}-p_T^{\mu_2}|)/(p_T^{\mu_1}+p_T^{\mu_2}) < 0.7.
$$ 

An additional cut with a similar scope as the initial $E_T$ cut 
was applied on the fraction of the
total transverse energy carried by the muon pair
$$ 
  0.1 < (p_T^{\mu_1}+p_T^{\mu_2})/E_T <
\begin{cases}
0.5 & \text{for }m^{\mu\mu} < 4~\gev \\
0.7 & \text{for }m^{\mu\mu}\geq 4~\gev.
\end{cases}
$$ 
The reason for the distinction of the two different dimuon mass
regions will be explained in Section \ref{sect:sigex}.
This $E_T$-fraction cut removes events where the hadronic activity is, 
respectively, very high
(false-muon background) 
or low (quarkonia and Bethe-Heitler).

Cosmic-ray muons were removed by discarding events with back-to-back muon
candidates 
and events in which the average calorimeter timing differs by more than 10~ns 
from
the nominal collision time. Large cosmic showers were removed using the 
BAC total energy and number of BAC muon segments. 

A sample of 4146 dimuon events was obtained using selection B.
Selection A retained about two thirds of these events.

\subsection{\label{sec-sel-iso}   Muon isolation}

Muons from semileptonic decays are usually not isolated, i.e. they are
normally 
accompanied by hadrons originating from the fragmentation and decay of the 
parent heavy quark and from other hadronic activity in the event. 
Hadronic activity in the detector was reconstructed using a combination of 
both track and calorimeter information \cite{thesis:briskin:1998}
referred to as energy-flow objects (EFOs).  
The difference in azimuth angle and pseudorapity, $\Delta \phi$ and 
$\Delta \eta$, was calculated between each EFO and 
each muon candidate in the event. 
The total transverse energy, $I_{1,2}$, deposited in a cone of
$\Delta R = \sqrt{\Delta \phi^2 + \Delta \eta^2}<1$ around each muon flight
direction was calculated by summing over all relevant EFOs, excluding the 
other muon. Since usually either both (beauty signal and open charm) or neither
(elastic $J/\psi$, Bethe-Heitler, etc.) of the muons arise from semileptonic 
decays, the quadratic sum $I^{\mu\mu}=\sqrt{I_1^2+I_2^2}$ of the two energy 
sums was found to yield the best sensitivity to distinguish between the two 
cases.

\section{Background and event simulation}

In order to measure the beauty signal,
several background contributions to the selected data sample were 
evaluated:
\begin{itemize}
\item the background from open charm decays not originating from beauty; 
\item the background from quarkonium states not originating from open beauty
      ($J/\psi$, $\psi^\prime$, $\Upsilon$, ...), 
      produced in elastic or inelastic collisions;
\item the background from Bethe-Heitler muon pair production;
\item the background from false muons.
\end{itemize}

Monte Carlo simulations of beauty and charm production were performed
using the generators {\sc Pythia} \cite{cpc:82:74} (for events with $Q^2<1~\gev^2$) and 
{\sc Rapgap} \cite{cpc:86:147} (for $Q^2>1~\gev^2$).
These simulations include the direct photon-gluon fusion process
($\gamma g \to Q \bar Q$, $Q=b,c$), flavour
excitation in the resolved photon and proton 
(e.g. $Q g \to Q g$, $\gamma Q \to Q g$), and hadron-like resolved photon
processes (e.g. $g g  \to Q \bar Q$).
Gluon splitting into heavy flavours ($g \to Q\bar Q$) in the initial or
final states of light-quark 
events was not included in the simulations;
this contribution is, however, expected to be small \cite{thesis:longhin:2004}.

Inelastic quarkonium production was simulated using {\sc Herwig} 
\cite{cpc:67:465}, while elastic quarkonia and Bethe-Heitler processes
were produced using several generators including {\sc Grape} \cite{Abe:2000cv}.

The ZEUS detector response, including the transformation of MC truth
level quantities into reconstructed quantities, was simulated in detail 
using a programme based on {\sc Geant} 3.13 \cite{GEANT}.
The detector simulation for beauty and charm events includes the 
simulation of both real and false muons.

Fake muons can be produced by hadron showers leaking from the back of the
 calorimeter
or by charged hadrons traversing the entire calorimeter without interaction.
In addition, low-momentum muons can originate from in-flight decays
of pions and kaons. Tracks reconstructed in the central tracker
may also be erroneously associated to a signal from a real muon in the muon 
chambers. A study \cite{thesis:longhin:2004}
 based on pions from $K^0$ decays, protons from $\Lambda$ decays, and
kaons from $\phi$ and $D^*$ decays, showed that
the detector simulation reproduced these backgrounds reasonably well. 
They will be collectively referred to as false muons.

Backgrounds from false muons in events not containing charm or beauty 
were not simulated. 
They were estimated from the data using the subtraction method described in 
Section \ref{sec:meas_principle}.

Since the muon range in dense material 
(effective momentum threshold) and the muon detector efficiencies
were imperfectly simulated,
corrections to the MC were determined \cite{thesis:bloch:2005-tmp-43fb5be7} 
using an independent data set consisting of isolated $\jpsi$ and 
Bethe-Heitler events.
Tabulated as a function of $p_T^\mu$ and $\eta^\mu$, these corrections  
were applied to MC events on an event-by-event basis.

\section{\label{sec-NLO} Theoretical predictions and uncertainties}

The MC programs described earlier, based on leading-order (LO) matrix
elements with the addition of parton showers (PS) to obtain higher-order 
topologies,
were used for the acceptance corrections. These programs are expected to 
describe the shapes of differential distributions, but not necessarily their
normalisation. 
For quantitative comparisons with QCD, next-to-leading-order (NLO) predictions
are used. 

QCD calculations in which $b$ quarks are treated as massless particles
\cite{bmassless1,*bmassless2,*bmassless3} are not applicable in the kinematic 
range relevant here.
Calculations based on CCFM parton-evolution schemes 
\cite{bkt1,*bkt2,*jung1,*jung2}, also called $k_T$ factorisation,
do not yet exist with full NLO implementation.
Fixed-order NLO calculations with massive $b$ quarks were therefore
chosen as the reference predictions.

The NLO FMNR program \cite{np:b412:225}
 evaluates parton-level cross sections for beauty 
in $\gamma p$ collisions (photoproduction) in the
fixed-order massive mode, for both pointlike and hadron-like photon
couplings to the heavy quarks.
The Weizs\"acker-Williams (WW) approximation 
with an effective $Q^2_{\rm max}$ cutoff of 25 GeV$^2$ ($\sim m_b^2$) 
\cite{zfp:88:612,*pr:45:729,*pl:b319:339-345}
was used to evaluate and include the DIS contribution to the 
cross sections, which is approximately 15\%.
This is in agreement with the DIS prediction from HVQDIS described below.

The parton-density functions used were CTEQ5M \cite{epj:c12:375}
 for the proton, and GRV-G-HO \cite{pr:d46:1973-1979}
 for the photon. The renormalisation and factorisation
scales $\mu$ were
chosen to be equal and parametrised by $\mu_0 = \sqrt{p_{T}^2+m_b^2}/2$, where
$p_{T}$ is the average transverse momentum of the two emerging $b$ quarks,
and $m_b=4.75$ GeV is the $b$-quark mass.
Such a scale choice is equivalent to the choice $\mu_0 = E_T/2$ or 
$\mu_0 = \sqrt{E_T^2 + Q^2}/2$ used in 
many jet measurements at the Tevatron \cite{Tevajet} and at HERA 
\cite{epj:c44:183}, and is expected to compensate somewhat for 
uncalculated higher-order contributions \cite{AGscales}.
An estimate of the theoretical uncertainty was obtained by simultaneously
varying $4.5 < m_b < 5.0$ GeV and $\mu_0/2 < \mu < 2\mu_0$ such that the
uncertainty was maximised. Typical uncertainties resulting from this procedure
(e.g. for the $b\bar b$ total cross section) are +60\% and $-$30\%.
Variations of the parton densities and the strong coupling 
parameter, $\Lambda_{\rm QCD}$, led to uncertainties which were much smaller
than the uncertainties related to mass and scale variations. They were 
therefore neglected.

Predictions for visible $\mu\mu$ final states were 
obtained by linking the FMNR parton-level predictions to the fragmentation 
and decay chain provided by {\sc Pythia} using the FMNR$\otimes$\-{\sc Pythia} 
interface \cite{PYTHIAFMNR,*thesis:nuncio:2008}. 
Additional parton showering was not 
applied\footnote{The {\sc MC@NLO} approach
\cite{jhep:06:029,*jhep:08:007}, which
allows the combination of NLO matrix elements with parton showers, is not yet
available for $ep$ interactions.}.
The branching ratios were corrected
to correspond to those obtained from the Particle Data Group (PDG) \cite{PDG},
as listed in Table \ref{tab:branchings}.
All other parameters, including those for fragmentation, and the 
procedure to obtain their uncertainties, were the same as in an 
earlier analysis \cite{epj:c50:1434}, and described elsewhere 
\cite{PYTHIAFMNR}.

The DIS part of the inclusive cross section is also calculated using 
the NLO predictions
from HVQDIS \cite{np:b452:109,*pl:b353:535,*pr:d57:2806}.
Only point-like contributions are included in
this prediction.
The parton density function used was CTEQ5F4 \cite{epj:c12:375}.
The renormalisation and factorisation
scales $\mu$ were
chosen to be equal and parametrised by $\mu_0 = \sqrt{p_T^2+m_b^2+Q^2}/2$.
The mass and scales were varied as for FMNR.
A scheme for the calculation of visible cross sections for correlated final
states, corresponding to the
FMNR$\otimes${\sc Pythia} interface described above,
was not available. Therefore, DIS cross-section comparisons are limited
to parton level, and the DIS contribution to the inclusive cross sections 
is included in the FMNR$\otimes${\sc Pythia} predictions via the WW 
approximation.

\section{Signal extraction}
\label{sect:sigex}

\paragraph*{Dimuon mass and charge separation}

As motivated in Section \ref{sec:meas_principle}, events were separated by the muon charges into like- and 
unlike-sign dimuon samples.
To differentiate between muon pairs from the cascade decay of 
\textsl{the same} $b$ quark and those from \textsl{different} $b$ 
quarks, the distributions were further separated depending on the dimuon 
invariant mass: 
low-mass dimuons with 
$m^{\mu\mu} < 4~\gev$, enriched in muons from the same $b$ quark, and 
high-mass dimuons with $m^{\mu\mu} > 4~\gev$,
containing dimuons originating from the decay of different $b$ quarks only. 
The dominant signal and background 
contributions to the four subsamples are summarised in 
Table~\ref{tab: mu_charge_mass_corrl}.

The resulting dimuon mass distributions for the low- and high-mass,
like- and unlike-sign subsamples for selection B are shown in Fig.~\ref{fig1}.
The MC distributions were in each case normalised to the data 
according to the
procedure described in the following subsections.
The high-mass region is already strongly beauty enriched, while
the low-mass region exhibits a significant contribution from $J/\psi$
production not originating from $B$ hadron\footnote{The term $B$ hadron 
includes $b$ baryons.} 
decays.
Such dimuon pairs tend to be isolated.

\paragraph*{\label{sec: isol_cuts}Dimuon isolation cuts}
To reduce this $J/\psi$ contribution, as well as corresponding contributions 
from
$\psi'$, $\Upsilon$ and Bethe-Heitler processes, a non-isolation requirement
was applied, based on the fact that muons from semileptonic decays are 
accompanied by hadrons from the same decay. 
The dimuon isolation variable $I^{\mu\mu}$, defined in 
Section \ref{sec-sel-iso}, was required to exceed 250 MeV, 
safely above the noise level of the CAL. This reduces the  
elastic quarkonium and Bethe-Heitler contributions to an almost negligible 
level.

Inelastic quarkonium and Bethe-Heitler events might pass the above cut because
hadrons from e.g. the proton 
remnant can accidentally end up in the isolation cone.  
For events in the $J/\psi$ and
$\psi^\prime$ mass peaks, where this background is largest, the cut was 
therefore raised to 2~GeV.

In summary, dimuons fulfilling the relation
$$ 
  I^{\mu\mu} \geq
\begin{cases}
2.0~\gev   & \text{for }m^{\mu\mu} \in [2.9,~3.25]\text{ GeV or }m^{\mu\mu} \in [3.6,~3.75]~\gev \\
0.25~\gev  & \text{otherwise}
\end{cases}
$$ 
are called non-isolated. This additional 
requirement is satisfied by 3500 events from selection B. 
The other events form a complementary
isolated background sample.

Figure~\ref{fig2} shows the muon $p_T$ and $\eta$ distributions for 
non-isolated
unlike-sign dimuon pairs, combining the low- and high-mass samples.
The remaining contribution from $J/\psi$, Bethe-Heitler, etc. processes
was normalised to the isolated background sample.
The charm contribution is small and was normalised to the charm signal in the
$D^* + \mu$ sample \cite{epj:c50:1434} as outlined in 
Section~\ref{sec:meas_principle}.
The different contributions to Fig.~\ref{fig2} are listed in Table \ref{tab:muons}.

\paragraph*{\label{sec-sig-extr}Signal evaluation}

The beauty signal and false-muon background were obtained using the 
procedure described in Section~\ref{sec:meas_principle}.
However, some further corrections are needed. 
Events from unlike-sign background sources, such as charm, which have been 
reconstructed as like-sign dimuon events due to false muons, are included both 
in the false-muon background estimation and in the MC samples. To avoid double 
counting, this (very small) contribution is subtracted from the MC samples. 
False muons in the beauty MC are considered as part of the signal. 

The signal extraction procedure according to 
Eq.\,(\ref{eq: n_beauty_calc}) relies on the 
unlike- and like-sign false-muon background 
contributions being equal.
A dedicated false-muon background study \cite{thesis:bloch:2005-tmp-43fb5be7} 
revealed a small residual excess of 
unlike-sign over like-sign background. This excess was corrected for using 
a multiplicative correction factor, $\alpha_{\rm corr}$, of 1.02 for the 
high-mass and 1.06 for the low-mass dimuon sample.
The beauty fraction was thus 
determined using a modified version of Eq. (\ref{eq: n_beauty_calc})
\begin{equation}
\label{eq: n_beauty_calc_corr}
N_{b\bar{b} \to \mu\mu} = \left( N_{\rm data}^{\rm u} -  \alpha_{\mathrm{corr}}\mal N_{\rm data}^{\rm l}
                       - \left( N_{\rm charm} + N_{\rm VM} + N_{\rm BH} \right)
                         \right)
                         \times \left(
                         \frac{N_{b\bar{b}}^{\rm u} + N_{b\bar{b}}^{\rm l} }
                              {N_{b\bar{b}}^{\rm u} - \alpha_{\mathrm{corr}}\mal N_{b\bar{b}}^{\rm l} }
                              \right)^{\rm MC} .
\end{equation}

A total of 1783 of the 3500 non-isolated events from selection B 
were found to originate
from beauty, corresponding to a beauty fraction of 51\%.

\section{\label{sec-syserr}Systematic uncertainties}
The main sources of systematic uncertainty for the measurement of 
visible cross sections are described in this section, in approximate 
order of importance.
The numbers in parentheses refer to the specific case of the inclusive visible 
cross section of Section \ref{sec:total}. 
Bin-by-bin uncertainties were evaluated for
the differential distributions where possible and appropriate. They are 
mostly similar to those derived for the inclusive visible cross section. 
Additional uncertainties introduced by the extrapolation to quark-level 
cross sections are discussed in Section \ref{sec:total}. 

\begin{itemize}
\item {\bf Muon efficiency correction.}
The muon efficiency, including the efficiency of the muon chambers and of 
the MUON-CTD matching, is known to about 7\% from a 
study based on an independent muon sample, and from the variance 
of the cross section when information from different muon detectors is used
independently \cite{thesis:bloch:2005-tmp-43fb5be7}. Conservatively, it is 
assumed to be fully correlated between the two muons ($\pm 15$\%).
\item {\bf Normalisation of charm background.}
The transfer of the normalisation of the charm contribution from the 
$D^*\mu$ analysis \cite{epj:c50:1434} to this analysis involves the following 
uncertainties: the statistical error of the fit of the charm 
contribution, $\pm$10\%; the inclusive branching ratio $c \to \mu$, $\pm$10\%;
the acceptance uncertainty due to charm fragmentation and decay spectra, 
$\pm$10\%; the fragmentation fraction $c \to D^{*\pm}$, $\pm$6\%; 
the branching 
ratio $D^{*\pm} \to K\pi\pi$, $\pm$3\%; and the use of all muon detectors 
(this analysis) versus the use of the barrel and rear muon chambers only 
\cite{epj:c50:1434}, $\pm$10\%. The influence of the correlation between
the fitted beauty and charm fractions in the $D^*\mu$ analysis 
\cite{epj:c50:1434} was found to be 
negligible. The normalisation of the charm contribution was varied by 21\%
according to the resulting combined uncertainty ($\pm 12$\%).
\item {\bf Normalisation of the Bethe-Heitler, $\pmb{J/\psi}$, etc.  backgrounds.}
The normalisation of the residual non-isolated contributions 
from Bethe-Heitler, charmonium,
and $\Upsilon$ production was varied by 
$\pm 50\%$ ($\pm 10$\%).
\item {\bf False-muon background.}
As a cross check for the determination of the false-muon background by 
the subtraction method, 
the probability of a reconstructed hadron to be misidentified as a 
muon was obtained from an inclusive dijet MC sample and tabulated as a function
of $p_T$ and $\eta$. Starting from 
a data sample with selection cuts identical to the present analysis, except 
that only one muon candidate was required, false-dimuon events were created 
by assuming a suitable additional hadron to be identified as a muon according
to this tabulated probability. After corrections for trigger efficiency,
and for the contribution from one false and one true muon obtained directly 
from the $b$ and $c$ MC, an independent background prediction was 
obtained \cite{thesis:samson:2008}. It
agreed very well in both normalisation and shape with that of 
the default subtraction method, thus confirming the method.
Since the uncertainty on this background is already implicitly contained
in the statistical error of the subtraction method, no explicit additional 
uncertainty was assigned.  
\item {\bf $\pmb{b}$ spectral shape uncertainty and $\pmb{b \bar b}$ correlations.} 
It was checked that the $b$-quark spectra from \pythia and \rapgap 
agree well with the corresponding spectra from the NLO predictions described 
below~\cite{thesis:longhin:2004}. 
To estimate the effect of variations of this shape, and of effects of 
variations of the $b \bar b$ correlations for different topologies on the 
efficiency, the efficiency was evaluated using the \pythia direct 
contribution only, or doubling the non-direct contributions (+4\%/$-12$\%).
\item {\bf $\pmb{B^0 \bar B^0}$ oscillations.}
The $B^0 \bar B^0$ oscillation parameter was varied by 8\%. This includes
the uncertainties of the mixing implementation in the MC models used 
($\pm 4$\%).
\item {\bf Other $\pmb{b}$ MC model uncertainties.}
This includes the
uncertainty of the procedure used to account for differences
of the branching ratios in the
signal MC and Table \ref{tab:branchings}, 
the uncertainty from $b$ fragmentation, and from the shape of the lepton 
spectrum from $b$ decays ($\pm$ 10\%).
\item {\bf $\pmb{c}$ spectral shape uncertainty and $\pmb{c \bar c}$ correlations.}
The direct and non-direct fractions for the charm background MC were varied 
in the same way as for beauty. The effect on the signal was small (+0\%/$-4$\%).
\item {\bf Trigger efficiency.} 
The uncertainty on the trigger efficiency was estimated by comparing
the efficiencies for muon and non-muon triggers in data and MC ($\pm 5$\%).    
\item {\bf Other uncertainties.}
Other uncertainties include the variation of the like-/unlike-sign ratio for 
the false-muon background by 3\% ($\pm 3$\%), the variation of the isolation
cuts by up to 500 MeV ($\pm 2$\%), the variation of the $E_T$ cut 
(energy scale) 
by 3\% ($\pm 2$\%), the variation of the $p_T^\mu$ cuts (magnetic field
uncertainty) by 0.3\% ($< 1$\%).
\end{itemize}   

The total systematic uncertainty (+25\%/$-28$\%) was obtained by adding the 
above contributions in quadrature. The uncertainties related to the 
background normalisation and the $b$ and $c$ spectral shape uncertainties 
were applied at a bin-by-bin level where relevant, while the others were 
added globally. 
A 2\% overall normalisation uncertainty 
associated with the luminosity measurement was not included.

\section{Total $\pmb{b \bar{b}}$ cross section}
\label{sec:total}

As a first step towards the extraction of the total cross section for 
$b \bar b$ production, a
visible cross section was
extracted for the maximum possible region in muon phase space allowed by the
preselection and the detector acceptance (selection B). The criterion 
that the muon detection probability should
be at least about 30\% per muon leads to the following phase space definition
at truth level:
\begin{itemize}
\item $-2.2< \eta < 2.5$ for both muons; 
\item $p_T > 1.5$ GeV for one of the two muons;
\item $p_T > 0.75$ GeV for the other muon, as well as
      $p > 1.8$ GeV for $\eta<0.6$, or ($p > 2.5$ GeV or $p_T > 1.5$ GeV) for
      $\eta>0.6$.
\end{itemize}
This cross-section definition refers to only one pair of muons per event. 
If there are more than two muons, muons directly originating from $B$ hadron 
decays are taken preferentially to form the pair.
A visible cross section for dimuon production
from beauty decays in this phase space
\begin{equation}
\label{eq:sigvis}
\sigma_{\rm vis} (ep \to b\bar b X \rightarrow \mu\mu X')
              = 55 \pm 7({\rm stat.})^{+14}_{-15} ({\rm syst.}) {\rm\ pb}
\end{equation}
was obtained.
This cross section includes muons from direct $B$-hadron decays, and indirect
decays via intermediate charm hadrons or $\tau$ leptons. The two muons can either
originate from the same $b$ quark, or from different quarks of the $b\bar b$
pair. Muonic decays of kaons, pions or other light hadrons were not included.

The measured cross section is 
larger than, but compatible 
with, the FMNR$\otimes${\sc Pythia} NLO 
prediction 
\begin{equation}
\label{eq:sigvisNLO}
\sigma_{\rm vis, NLO}(ep \to e b \bar{b} X \to e \mu\mu X) = 33^{+18}_{-8} \rm{(NLO)}^{+5}_{-3}\rm{(frag. \oplus br.)}~\rm{pb},       
\end{equation}
where the first error refers to the uncertainties of the FMNR parton-level 
calculation, and the second error refers to the uncertainties
related to fragmentation and decay.

The visible cross section was then translated into the total
cross section for beauty production.  
The effective branching fraction of a
$b\bar b$ pair into at least two muons is 6.3\% 
\cite{cpc:82:74,PDG}.
The probability (acceptance) for such a muon pair to be in the kinematic 
range of the 
measured visible cross section, evaluated from the beauty MC sample, is 
about 6\% on average. 
Defining $p_{T,b}^{\rm max}$ as 
the maximum of the two $b$-quark transverse momenta after parton showering, and
$|\zeta_b|^{\rm min}$ as the minimum of the modulus of the rapidity
(not pseudorapidity) of the two quarks, 
this probability ranges from 3\% for 
$p_{T,b}^{\rm max} = 0$ GeV  to 9\%
at\footnote{At even larger $p_{T}^b$ the 
acceptance rises further, but the fraction of events is small.}
$p_{T,b}^{\rm max} = 10$ GeV, 
 for $|\zeta_b|^{\rm min} <2$. The acceptance is almost independent of rapidity
 within this rapidity range, which covers 90\% of the total $b\bar b$ phase 
space. It drops sharply at larger rapidities. 
Thus, only 10\% of the total beauty 
contribution in the region $|\zeta_b|^{\rm min} >2$ remains unmeasured.
The small dependence of the acceptance on the transverse momenta of the 
$b$ quarks is due to the low muon-momentum threshold, in combination with the 
large $b$-quark mass and the three-body decay kinematics. 
Sensitivity down to $p_{T}^b = 0$ GeV is obtained.

In summary, the combined probability for a $b\bar b$ pair to yield
a muon pair in the visible kinematic range (6.3\%$\times$6\%=0.38\% on average)
 is quite
small, but varies by less than a factor 3 over 90\% of the total phase space.
Furthermore, it is almost entirely determined by quantities measured\cite{PDG} 
with good precision at $e^+e^-$ colliders. 
These include the branching fractions listed in Table \ref{tab:branchings}, 
the $b$-fragmentation functions, and the $B$ hadron $\to \mu X$ decay spectra.
It was checked that all of these are well reproduced by the MC after the 
application of branching-ratio corrections.
The $b$-quark $p_T$ and rapidity spectra predicted by the {\sc Pythia} and
{\sc Rapgap} generators were found to agree with those from FMNR and 
HVQDIS to within 15\% \cite{thesis:longhin:2004}. Furthermore,
the quasi-uniformity of the acceptance explained above implies that the 
dependence on details of the simulation of the $b\bar b$ topology is rather 
weak. The MC can therefore 
safely be used for the extraction of the total cross section for 
beauty production.

The normalisation of the
{\sc Pythia} + {\sc Rapgap} MC prediction for the beauty contribution had to
be scaled up
by a factor 1.84 to agree with the dimuon data. 
Applying this measured scale factor to the total {\sc Pythia} and
{\sc Rapgap} cross sections, the total cross section for
$b\bar b$ pair
production in $ep$ collisions at HERA for $\sqrt{s}=318$ GeV was determined to be
\begin{equation}
\label{eq:sigq}
 \sigma_{\rm tot}(ep \to b\bar b X) =
   13.9 \pm 1.5({\rm stat.}) ^{+4.0}_{-4.3}({\rm syst.}) {\rm\ nb} ,
\end{equation}
where the first uncertainty is statistical and the second systematic.
In addition to the uncertainties described in Section \ref{sec-syserr}, this includes 
an error of 5\% from the uncertainties of the spectral shape mentioned 
above, and an error of 6\% from the variation of the branching ratios, 
added in quadrature.

The total cross section predicted by next-to-leading-order QCD
calculations
was obtained in the massive approach by adding the predictions from
FMNR \cite{np:b412:225,*np:b454:3-24,*frixione3} and HVQDIS~\cite{np:b452:109,*pl:b353:535}
for $Q^2$ less than or larger than 1 GeV$^2$, respectively. 
The resulting cross section for $\sqrt{s} = 318$ GeV
$$ 
\sigma_{\rm tot}^{\rm NLO}(ep \to b\bar b X)
   = 7.5 ^{+4.5}_{-2.1} {\rm\ nb} 
$$ 
is a factor 1.8 lower than the measured value, although compatible
within the large uncertainties.
The corresponding cross section from FMNR only using the Weizs\"acker-Williams
approximation to estimate the DIS contribution is
\begin{equation}
\label{eq:sigqWW}
 \sigma_{\rm tot}^{\rm WW}(ep \to b\bar b X)
   = 7.8^{+4.9}_{-2.3} {\rm\ nb}, 
\end{equation}
in agreement with the more exact FMNR+HVQDIS calculation.

The fact that the comparisons between data and theory 
yield the same ratio at the visible level 
(Eqs. (\ref{eq:sigvis})/(\ref{eq:sigvisNLO})):  
$$ 
 R_{\rm vis}^{\rm data/NLO} = 1.7^{+0.7}_{-1.1}\ ;
$$ 
and at quark level (Eqs. (\ref{eq:sigq})/(\ref{eq:sigqWW})): 
$$ 
R_{b}^{\rm data/NLO} = 1.8^{+0.8}_{-1.3}
$$ 
confirms the validity of the 
extrapolation procedure used.

Figure~\ref{fig3} shows a comparison of the measured total cross section to 
cross sections and theoretical predictions from the $D^*+\mu$ final state 
obtained by ZEUS in
earlier measurements \cite{epj:c50:1434}. Although not fully inclusive, 
these measurements are closest in phase space to the measurement presented 
here.
Qualitatively, they show the same trend of the cross sections 
being higher than, but consistent with, the corresponding QCD predictions.
The somewhat larger deviations reported in 
similar $D^*+\mu$ measurements by H1 \cite{pl:b621:56-71} are not supported.

\section{Differential cross sections and $\pmb{b \bar{b}}$ correlations}

Selection A was used for the measurement of visible differential 
cross sections because a uniform kinematic 
acceptance is more relevant than maximal phase-space coverage.
Correspondingly, at truth level, the phase space was restricted to:
\begin{itemize}
\item $p_{T}^{\mu} > 1.5$ GeV for both muons
\item $-2.2 < \eta^\mu < 2.5$ .
\end{itemize}
The backgrounds were again normalised as described in 
Section \ref{sect:sigex}.
The signal-extraction procedure was the same as for
the inclusive visible cross section, except for being applied
bin by bin.
Bin-dependent systematic uncertainties were calculated wherever possible.
The resulting cross sections for the differential $p_T^\mu$ and $\eta^\mu$
spectra are shown in Figs.~\ref{fig4} and \ref{fig5}.
Very good agreement is observed with the
{\sc Pythia}+{\sc Rapgap} predictions scaled by the same factor 1.84
that was measured for the total cross section.
Apart from the normalisation, the leading-order plus parton-shower (LO+PS)
approach yields a good description of the corresponding physics
processes within the entire accessible phase space.
This confirms the applicability of these MC models for acceptance 
calculations.

A comparison of the measured cross sections to the absolute FMNR$\otimes${\sc Pythia}
NLO QCD predictions is also shown in Figs. \ref{fig4} and \ref{fig5}. 
Again, good agreement in 
shape is observed, with a tendency to underestimate the data normalisation 
consistent with the observations from the total cross section.
A potential trend for increasing data/theory deviations towards low $p_T$ and/or 
high $\eta$, 
suggested by some previous measurements \cite{epj:c41:453,pl:b599:173}, is not supported.

To provide a more detailed look at the correlations between the two $b$ quarks,
the reconstructed dimuon mass range was restricted\footnote{While the mass separation value of 4 GeV described earlier was optimised such that all dimuons 
from the same $b$ quark contribute to the low-mass sample, including dimuons from 
$b \to \psi\prime$ decays, the value 3.25 GeV was chosen to optimise the 
separation power for dimuons from same and different $b$ quarks.}
to $m^{\mu\mu} > 3.25$ GeV.
This reduced the contribution of dimuons from the same quark to an almost
negligible level. The corresponding data distribution for $\Delta \phi$
between the two muons is shown in Fig.~\ref{fig6}.
Figure~\ref{fig7} shows the resulting differential cross section, where the 
mass cut was replaced by the requirement that the two muons originate from 
different $b$ quarks.
The distribution is well
described by the FMNR$\otimes${\sc Pythia}
NLO QCD predictions within the large uncertainties resulting from the
subtraction method (Eq. (\ref{eq: n_beauty_calc_corr})).

\section{Hadron- and parton-level cross-sections}
\label{sec-hadron}

In order to compare to previous ZEUS results
using other final states 
\cite{pr:d70:012008,pl:b599:173,*epj:c18:625}, expressed in terms of 
parton-level cross sections\footnote{H1 results have not been 
published in this form.} differential in $p_{T}^b$, similar cross sections 
were also extracted. 

The first step was the extraction of visible cross sections for $B$ hadrons in 
different $p_T$ ranges. For this purpose, the data sample used for the 
measurement of the total beauty cross section (selection B) was split into two 
subsamples, 
with $m^{\mu\mu} > 3.25$ GeV and $m^{\mu\mu} < 3.25$ GeV. As motivated in the 
previous section, the $m^{\mu\mu} > 3.25$ sample is dominated by muons from 
different $b$ quarks, with correlations between the two quarks
which are reasonably understood. Thus, two measured $B$ hadrons are present 
in each event.
To estimate their transverse momenta, the quantitity
$$ 
E_T^{\rm vis} = p_T^\mu + I^\mu 
$$ 
is evaluated for each muon, where $I^\mu$ is the cone transverse energy 
described in Section \ref{sec-sel-iso}.
This variable is found to be strongly correlated to the parent $B$ hadron 
transverse momentum at high 
$p_T$, where the additional energy from $b$-quark fragmentation to the 
$B$ hadron compensates the loss due to the unreconstructed neutrino from 
the semileptonic decay. 
At $p_T \lesssim m_b$, this correlation is diluted by the effect of the 
$B$-hadron mass and the corresponding decay kinematics.
Figure \ref{fig:ptBbins}(a) shows the expected $B$-hadron $p_T$ spectra for 
three bins in 
$E_T^{\rm vis}$, 
$0<E_T^{\rm vis}<5$ GeV, $5<E_T^{\rm vis}<10$ GeV, and $10<E_T^{\rm vis}<40$ GeV. 
Reasonably distinct $B$-hadron $p_T$ regions are probed.
The corresponding visible cross sections are shown in Fig. \ref{fig:ptB}(a).

A similar procedure was applied to the subsample with  $m^{\mu\mu} < 3.25$ GeV.
In this sample, the muons originate mainly from the same $b$ quark, therefore
only one $B$ hadron has been measured. Due to branching ratios and decay 
kinematics,
the cross section is smaller, but the absence of like-sign muon pairs from 
the same $b$ quark leads to a smaller uncertainty from the subtraction method.
Therefore, the precision of
the measurement is comparable to that from the high-mass 
region. Furthermore, the subtraction method reduces the influence of the 
residual contribution of muons from different $b$ quarks. Thus, the measured 
cross sections are almost completely insensitive to $b \bar b$ correlations.

The $E_T^{\rm vis}$ variable is redefined to 
$$ 
E_T^{\rm vis} = p_T^{\mu\mu} + I^\mu_{\rm high} 
$$ 
where  $p_T^{\mu\mu}$ is the transverse momentum of the dimuon system
added vectorially, and $I^\mu_{\rm high}$ is the isolation of the 
higher $p_T$ muon only, to avoid double counting.
The correlations to the $B$ hadron $p_T$ are similar to the high-mass
case (Fig. \ref{fig:ptBbins}(b)), enabling them to be combined later on.
The resulting visible $B$-hadron cross sections are shown in Fig. 
\ref{fig:ptB}(b).

For both subsamples, agreement is found with the  
FMNR$\otimes${\sc Pythia} predictions, consistent with the conclusions
obtained earlier.

The second step is to extrapolate these cross sections to $b$-quark
level. For comparison with previous measurements, 
the cross sections were restricted to photoproduction. 
Each of the $B$-hadron visible cross sections is translated into 
a differential cross section
$\frac{d\sigma}{dp_{T}^b}$ in the pseudorapidity range $|\eta_b| < 2$
\cite{pr:d70:012008}
 with photon virtuality $Q^2 < 1$ GeV$^2$ and inelasticity 
$0.2 < y < 0.8$, using the FMNR$\otimes${\sc Pythia} predictions.  
Each cross section is quoted at the mean $p_{T}^b$ value for events
satisfying the cuts for the corresponding $E_T^{\rm vis}$ bin.
The results are shown in Fig. \ref{fig:ptbbins}. 

The cross sections derived from the low- and high-mass subsamples
(same and different $b$ quarks) are in agreement, and were 
combined to give a single cross section for each $p_T^b$ value. 
The maximum possible correlation of the systematic errors is assumed
for this combination.

The resulting combined cross sections are compared to theory and 
previous 
measurements in Fig.\,\ref{fig9}. 
They are consistent with these previous measurements, and extend the 
measured range to lower $p_T^b$.
Predictions at NLO \cite{np:b412:225} and predictions from a LO 
$k_T$-factorisation approach \cite{bkt1} yield an equally good description 
of the data.

\section{Conclusions}

The total cross section for beauty production in $ep$ collisions at 
$\sqrt{s} = 318$ GeV has been measured for the first time
using an analysis technique based on the
detection of two muons, mainly from semileptonic beauty decay.
The almost complete phase-space coverage combined with the weak dependence 
on details
of the $b\bar b$ event topology allowed a reliable extraction of the total
beauty production cross section, with acceptance down to $p_{T}^b=0$ GeV,
and a direct comparison to NLO QCD predictions.
The predictions are lower than the observed cross sections, but compatible 
within the uncertainties.
Differential cross sections in $p_T^\mu$, $\eta^\mu$, and
$\Delta\phi^{\mu\mu}$
were also measured. 
Shapes predicted
by Monte Carlo models incorporating leading-order matrix elements followed by
parton showers agree well with the data.
NLO QCD predictions agree in shape with both the data and the LO+PS predictions,
but are again somewhat lower than the data, in agreement with the observation 
from the total cross section. 
The angular correlations between final-state muons from different $b$ quarks, 
reflecting the correlations between these parent quarks, are described by the
NLO QCD predictions. 
Measurements of cross sections for muon pairs from the same or from 
different $B$ hadrons yield similar and compatible results.
A comparison with previous measurements through the extrapolation to differential
cross sections at $b$-quark level shows reasonable agreement, and extends these
measurements down to lower $p_T^b$.

\section*{\label{sec-ackno}Acknowledgements}
We thank the DESY Directorate for their strong support and encouragement.
The remarkable achievements of the HERA machine group were essential for the
 successful completion of this work and are greatly appreciated.
 We are grateful for the support of the DESY computing and network services.
The design, construction and installation of the ZEUS detector have been
made possible owing to the ingenuity and effort of many people who are not
listed as authors.
It is also a pleasure to thank S. Frixione
for help with the theoretical predictions.

\vfill\eject

{
\def\bibname{\Large\bf References}
\def\refname{\Large\bf References}
\pagestyle{plain}
\ifzeusbst
  \bibliographystyle{./BiBTeX/bst/l4z_default}
\fi
\ifzdrftbst
  \bibliographystyle{./BiBTeX/bst/l4z_draft}
\fi
\ifzbstepj
  \bibliographystyle{./BiBTeX/bst/l4z_epj}
\fi
\ifzbstnp
  \bibliographystyle{./BiBTeX/bst/l4z_np}
\fi
\ifzbstpl
  \bibliographystyle{./BiBTeX/bst/l4z_pl}
\fi
{\raggedright
\bibliography{./BiBTeX/user/syn.bib,%
              ./BiBTeX/bib/l4z_articles.bib,%
              ./BiBTeX/bib/l4z_books.bib,%
              ./BiBTeX/bib/l4z_conferences.bib,%
              ./BiBTeX/bib/l4z_h1.bib,%
              ./BiBTeX/bib/l4z_misc.bib,%
              ./BiBTeX/bib/l4z_old.bib,%
              ./BiBTeX/bib/l4z_preprints.bib,%
              ./BiBTeX/bib/l4z_replaced.bib,%
              ./BiBTeX/bib/l4z_temporary.bib,%
              ./BiBTeX/bib/l4z_zeus.bib}}
}
\vfill\eject

\begin{table}[hbt]
\begin{center}
\begin{tabular}{|c|r|}
\hline
channel & effective branching fraction w/o $B^0 \bar B^0$ mixing \\
\hline
$b \to \mu^-$ direct & $10.95 \pm 0.27$ \% \hspace{2.5cm} \\
 $b \to \mu^{+}$ indirect & $8.27 \pm 0.40$ \% \hspace{2.5cm} \\
 $b \to \mu^{-}$ indirect & $2.21 \pm 0.50$ \% \hspace{2.5cm} \\
\hline
all $b \to \mu^{\pm} $ & $21.43 \pm 0.70$ \% \hspace{2.5cm} \\
\hline
$ b \bar b \to \mu^{\pm} \mu^{\mp} $ (diff. $b$s) & $2.42 \pm 0.17$ \% \hspace{2.5cm} \\
$ b \bar b \to \mu^{\pm} \mu^{\pm} $ (diff. $b$s) & $2.18 \pm 0.14$ \% \hspace{2.5cm} \\
\hline
$b \to \mu^{+} \mu^- $ all & $2.40 \pm 0.16$ \%  \hspace{2.5cm} \\
\hline
\end{tabular}
\end{center}
\caption{
Effective branching fractions used for cross-section 
determinations. 
The indirect contributions include cascade decays into muons via charm, 
anticharm, $\tau^\pm$ and $J/\psi$. The additional effect of 
$B^0 \bar B^0$ mixing ($\chi = 0.1283 \pm 0.0076$) is not included.}
\label{tab:branchings}
\end{table}

\begin{table}
  \begin{center}
    \begin{tabular}[htbp]{|c||c|c|}
    \hline
                  & unlike-sign $\pm / \mp$             & like-sign $++/--$ \\
      \hline       \hline
                  &                                       &\\
      low inv. mass & {\bf muons from same b},            & {\bf false-muon background}, \\
$m_{\mu\mu}<4~\gev$ & muons from $J/\psi$, $\psi'$,       & and small contribution of  \\
                    & and false-muon background          & muons from different b  \\
                  &                                       &\\
      \cline{1-3}
                  &                                       &\\
     high inv. mass & {\bf muons from different b},  & {\bf muons from different b} \\
$m_{\mu\mu}>4~\gev$ & muons from $c\bar c$, $\Upsilon$, BH, & and false-muon background\\
                    & and false-muon background           &                          \\
                  &                                       &\\
    \hline
    \end{tabular}
  \end{center}
  \caption[~Dimuon mass and charge correlations]{
    \label{tab: mu_charge_mass_corrl}
    Classification of events using dimuon mass and charge correlations.
    The main contributions to each class are listed; 
    the most relevant is indicated in bold face.}
\end{table}

\begin{table}[hbt]
\begin{center}
\begin{tabular}{|c|c|}
\hline
process & muon candidates \\
\hline
beauty           & 2382 \\
charm            &  629 \\
quarkonia and BH &  281 \\
false muon       & 1281 \\
\hline
data             & 4574 \\
\hline
\end{tabular}
\end{center}
\caption{
Number of muon candidates contributing to Fig. 2: 
unlike-sign non-isolated dimuons. 4574 muons correspond to 2287 events.}
\label{tab:muons}
\end{table}

\setlength{\unitlength}{1mm}
\begin{figure}[p]
  \begin{center}
    \begin{picture}(160,139)
      \put(0,-3){\epsfig{file=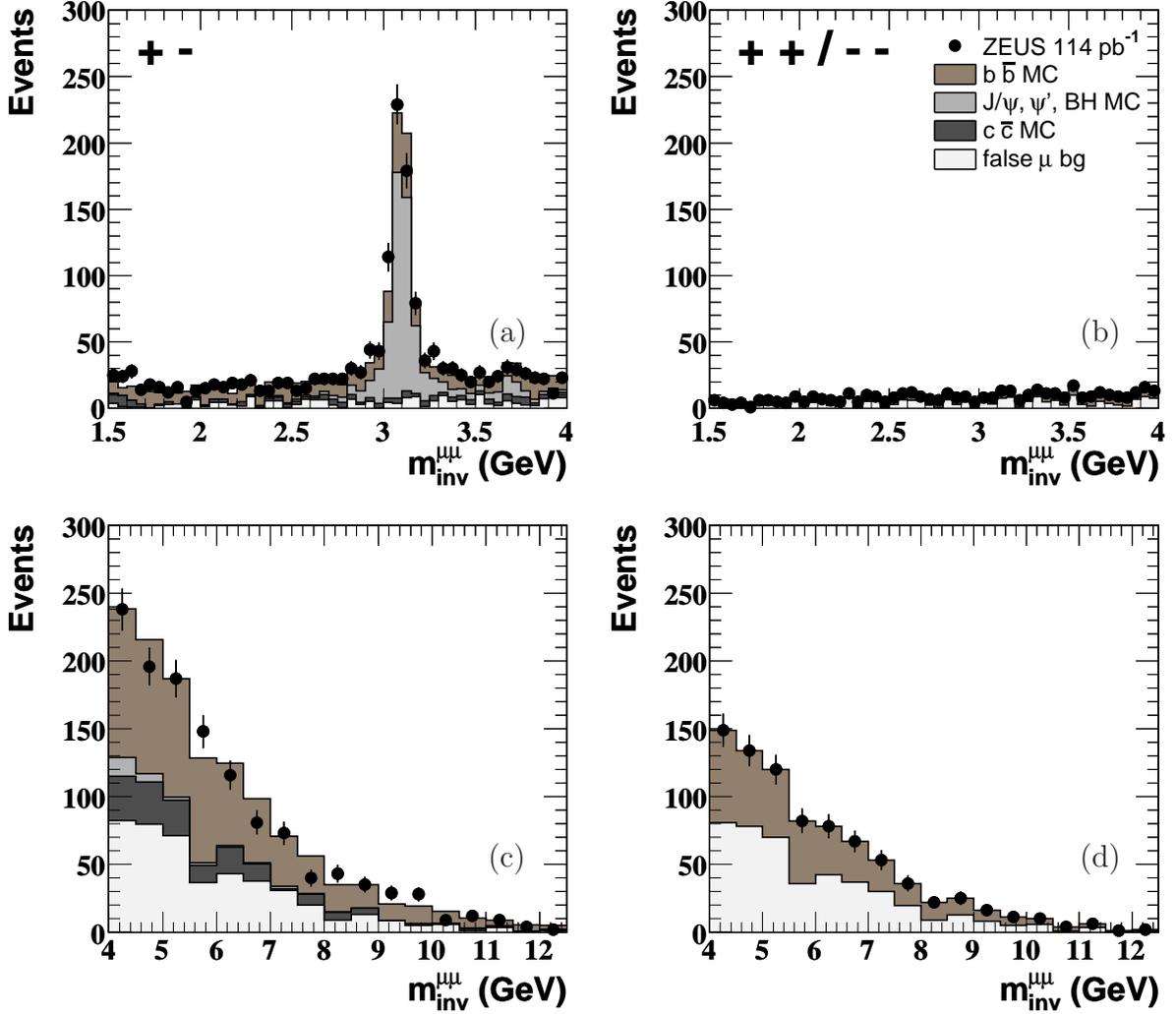, width=16cm, clip}}
      \put(67,88){(a)} 
      \put(147,88){(b)} 
      \put(67,17){(c)} 
      \put(147,17){(d)} 
    \end{picture}
  \end{center}
  \caption{
    Dimuon mass distributions 
    of unlike-sign dimuon pairs from selection B (see text) 
    in the (a) low-mass and (c) high-mass 
    subsamples, as well as like-sign dimuon pairs in the (b) low-mass and 
    (d) high-mass subsamples. The same vertical scale has been chosen for the 
    like- and unlike-sign subsamples, with different bin sizes for the high- 
    and low-mass regions.  
    The expected contributions from different processes are
    also shown. The false-muon background was obtained from the data using
    the subtraction method described in Section \ref{sec:meas_principle}.
    Due to this method, the total prediction for like-sign pairs agrees with 
    the data by definition. 
  }
  \label{fig1}
\end{figure}

\begin{figure}[p]
  \begin{center}
    \begin{picture}(160,70)
      \put(0,-3){\epsfig{file=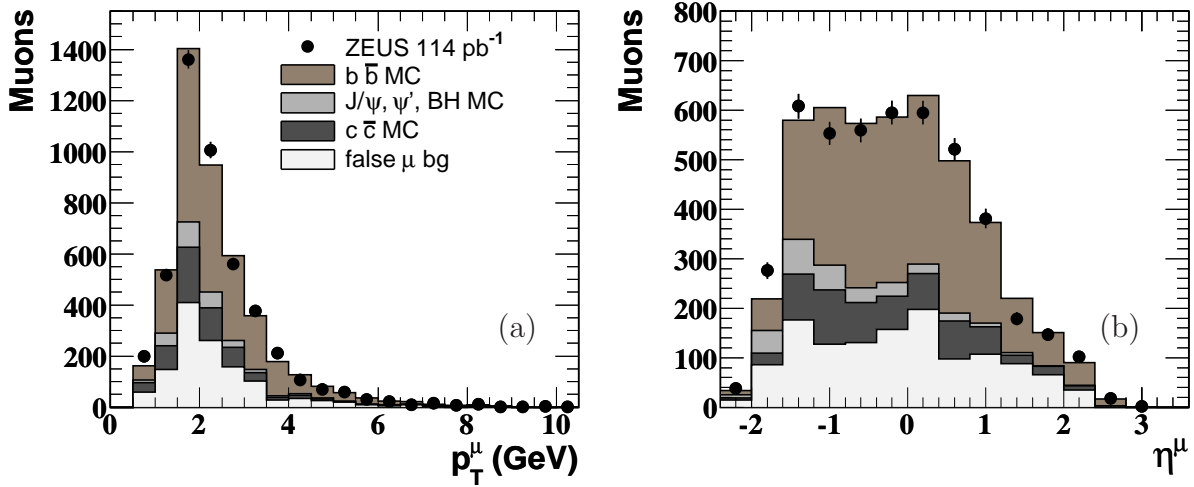, width=15.9cm, clip}}
      \put(67,17){(a)} 
      \put(147,17){(b)} 
    \end{picture}
  \end{center}
  \caption{
    (a) Muon transverse momentum and (b) muon pseudorapidity distribution from 
    both high- and low-mass dimuon pairs in the non-isolated unlike-sign 
    sample.
    Two muons are entered for each event.
    The expected contributions from different processes are also shown. 
    Due to the subtraction method, the statistical error of the prediction
    for  
    the false-muon background is comparable in absolute size to that of 
    the data. 
  }
  \label{fig2}
\end{figure}

\begin{figure}[p]
  \begin{center}
  \vskip 5 cm
  \resizebox*{0.85\textwidth}{!}{
    \includegraphics{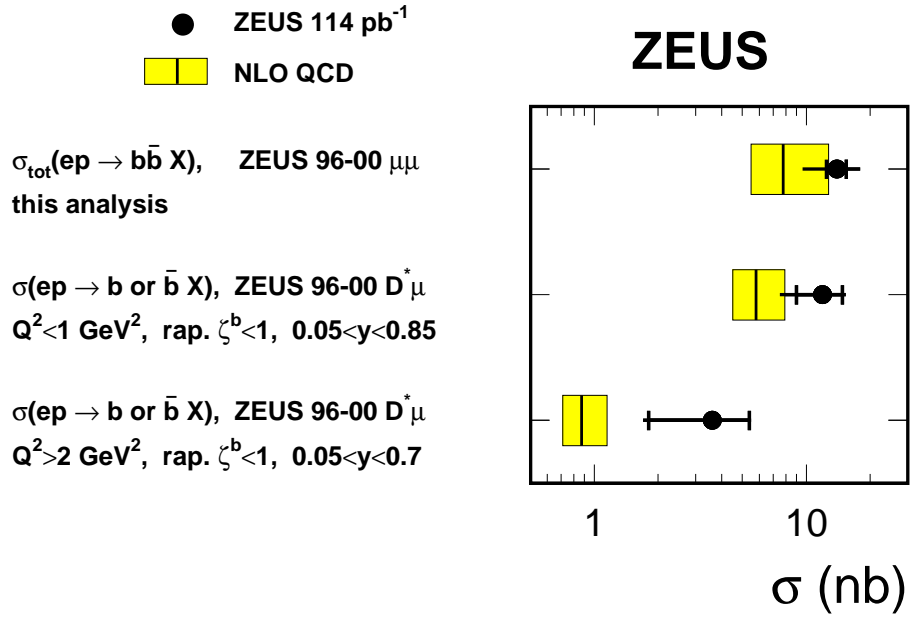}
  }
  \end{center}
  \caption{
    Comparison of measured cross sections to NLO QCD predictions. 
    The $b\bar{b}$ cross section from this analysis (top) is  
    compared to both measured and predicted $b$~or~$\bar{b}$ cross sections
    obtained in the ZEUS $D^*\mu$ analysis \protect\cite{epj:c50:1434} 
    for the photoproduction 
    regime (middle line)
    and DIS (lower line).
    The NLO calculations in the $D^*\mu$ analysis used a slightly different 
    set of parameters. Using the parameters detailed in Section \ref{sec-NLO},
    the central value of the photoproduction cross-section prediction would
    increase by about 20\%.  
  } 
  \label{fig3}
  \vskip 5 cm
\end{figure}

\begin{figure}[htbp]
  \begin{center}
  \resizebox*{0.5\textwidth}{!}{
    \includegraphics{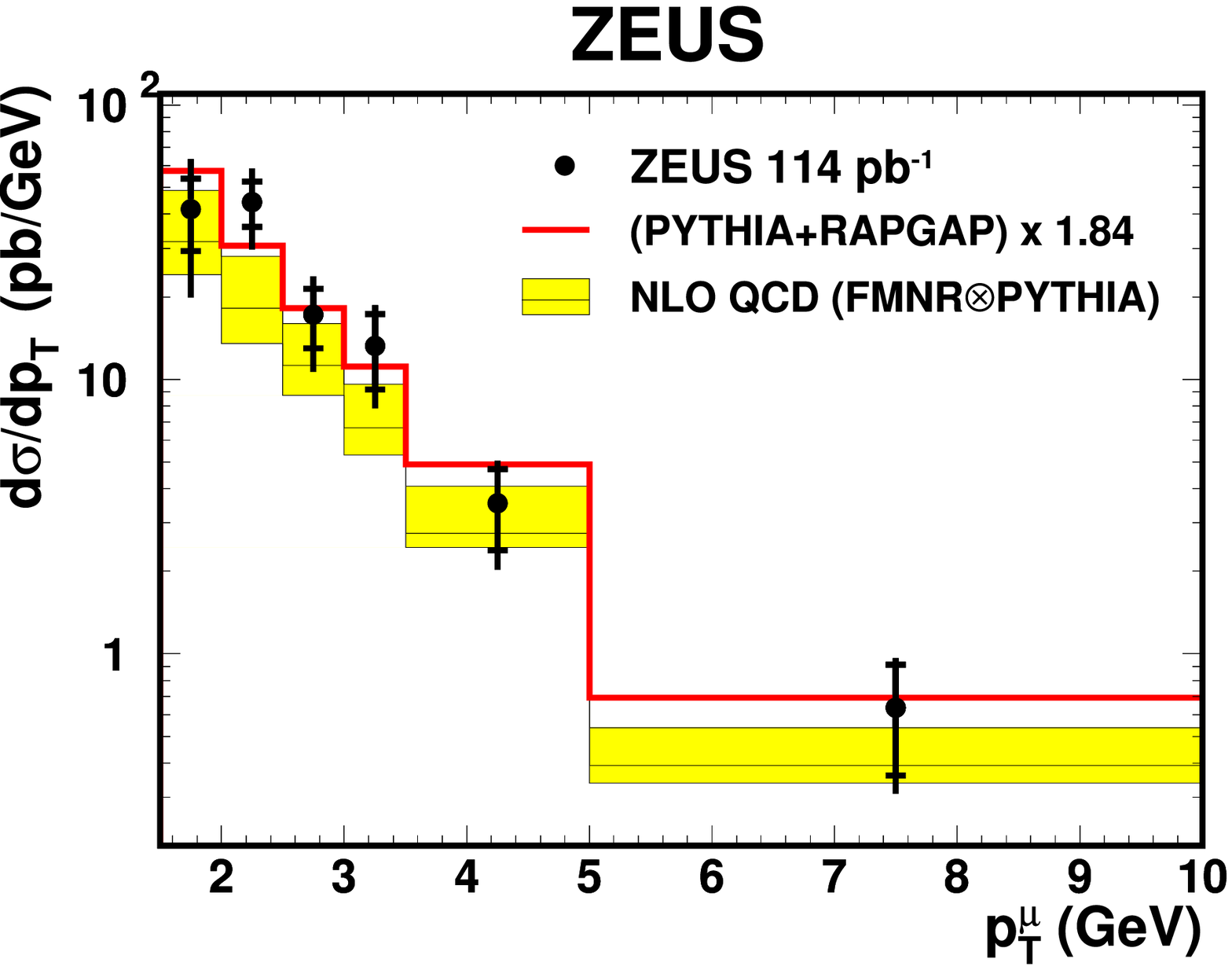}
  }
  \vskip 1cm
  \end{center}
  \caption{
    Cross-section $d\sigma/dp_T^\mu$ for muons from b decays in 
    dimuon events with $p_T^\mu>1.5$ GeV and $-2.2<\eta^\mu<2.5$ for both 
    muons. Two muons contribute for each event.
    The data (solid dots) are 
    compared to the scaled sum of the predictions by the LO+PS generators 
    {\sc Pythia} and {\sc Rapgap} (histogram)
    and to the NLO QCD predictions from {\rm FMNR}$\otimes${\sc Pythia} (band).
  } 
  \label{fig4}
\end{figure}

\begin{figure}[htbp]
  \begin{center}
  \vskip 1cm
  \resizebox*{0.5\textwidth}{!}{
    \includegraphics{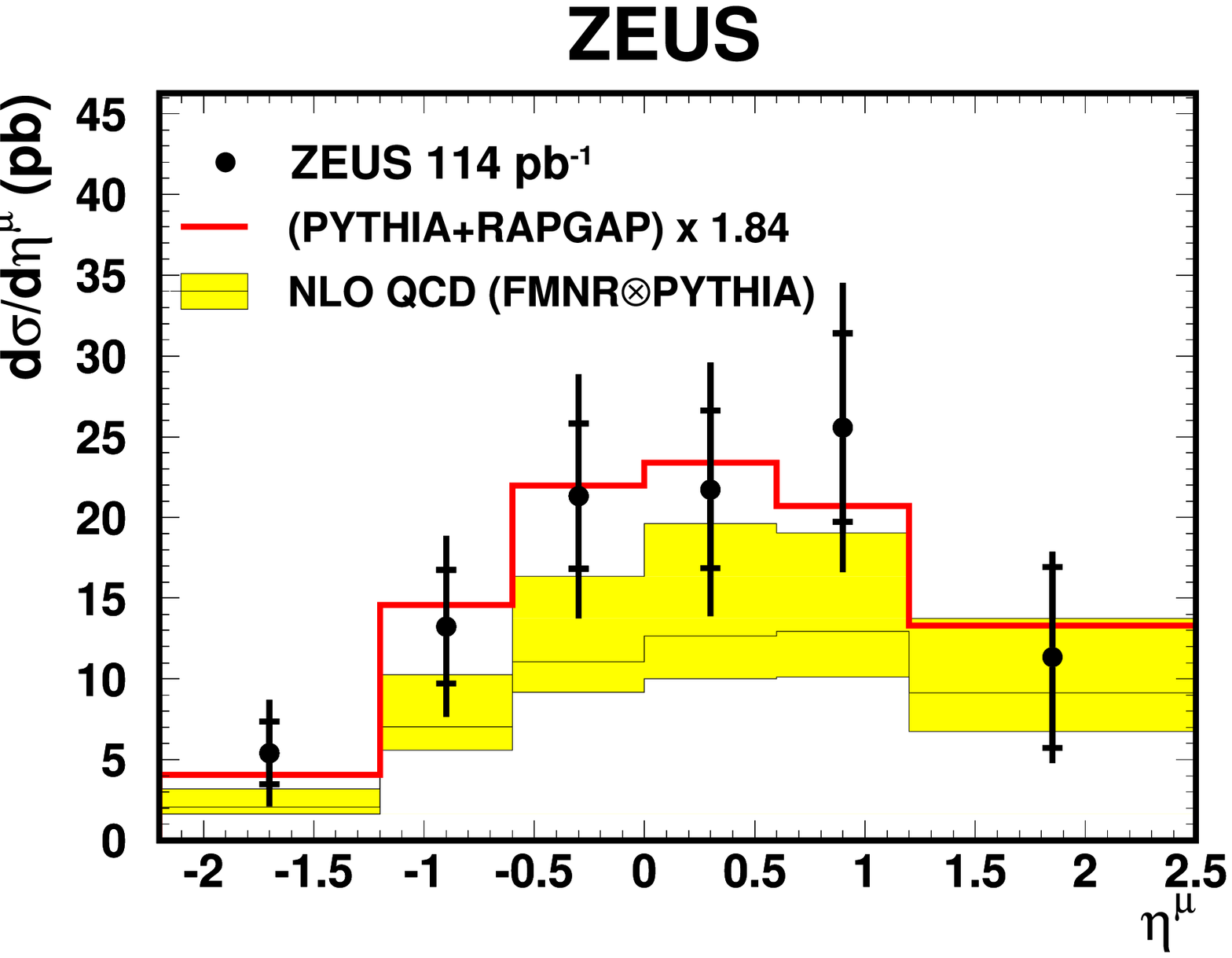}
  }
  \vskip 1cm
  \end{center}
  \caption{
    Cross-section $d\sigma/d\eta^\mu$ for muons from b decays in 
    dimuon events with $p_T^\mu>1.5$ GeV and $-2.2<\eta^\mu<2.5$ for both 
    muons. Two muons contribute for each event. 
    The data (solid dots) are 
    compared to the scaled sum of the predictions by the LO+PS generators 
    {\sc Pythia} and {\sc Rapgap} (histogram)
    and to the NLO QCD predictions from {\rm FMNR}$\otimes${\sc Pythia} (band).
  } 
  \label{fig5}
\end{figure}

\begin{figure}[htbp]
  \begin{center}
  \resizebox*{0.5\textwidth}{!}{
    \includegraphics{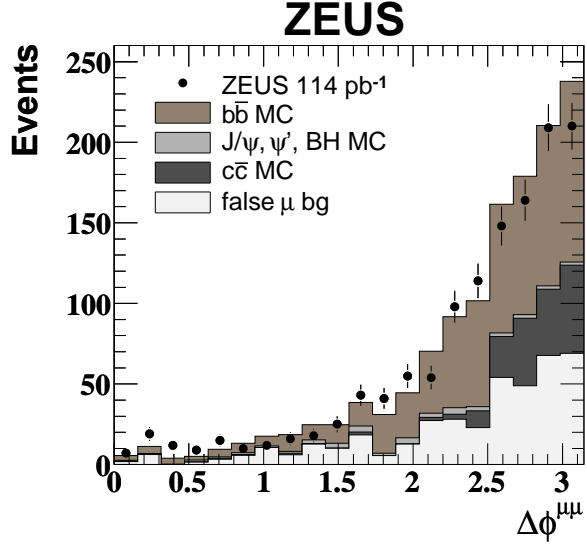}
  }
  \end{center}
  \caption{
    Distribution of the azimuthal distance $\Delta\phi$ between the two muons
    in dimuon events with $p_T^\mu>1.5$ GeV 
    for both muons, and $m^{\mu\mu} > 3.25$ GeV.
    The expected contributions from different processes are
    also shown. 
    Due to the subtraction method, the statistical error of the prediction
    for the false muon background is comparable in absolute size 
    to that of the data. 
  } 
  \label{fig6}
\end{figure}

\begin{figure}[htbp]
  \begin{center}
  \vskip 1cm
  \resizebox*{0.5\textwidth}{!}{
    \includegraphics{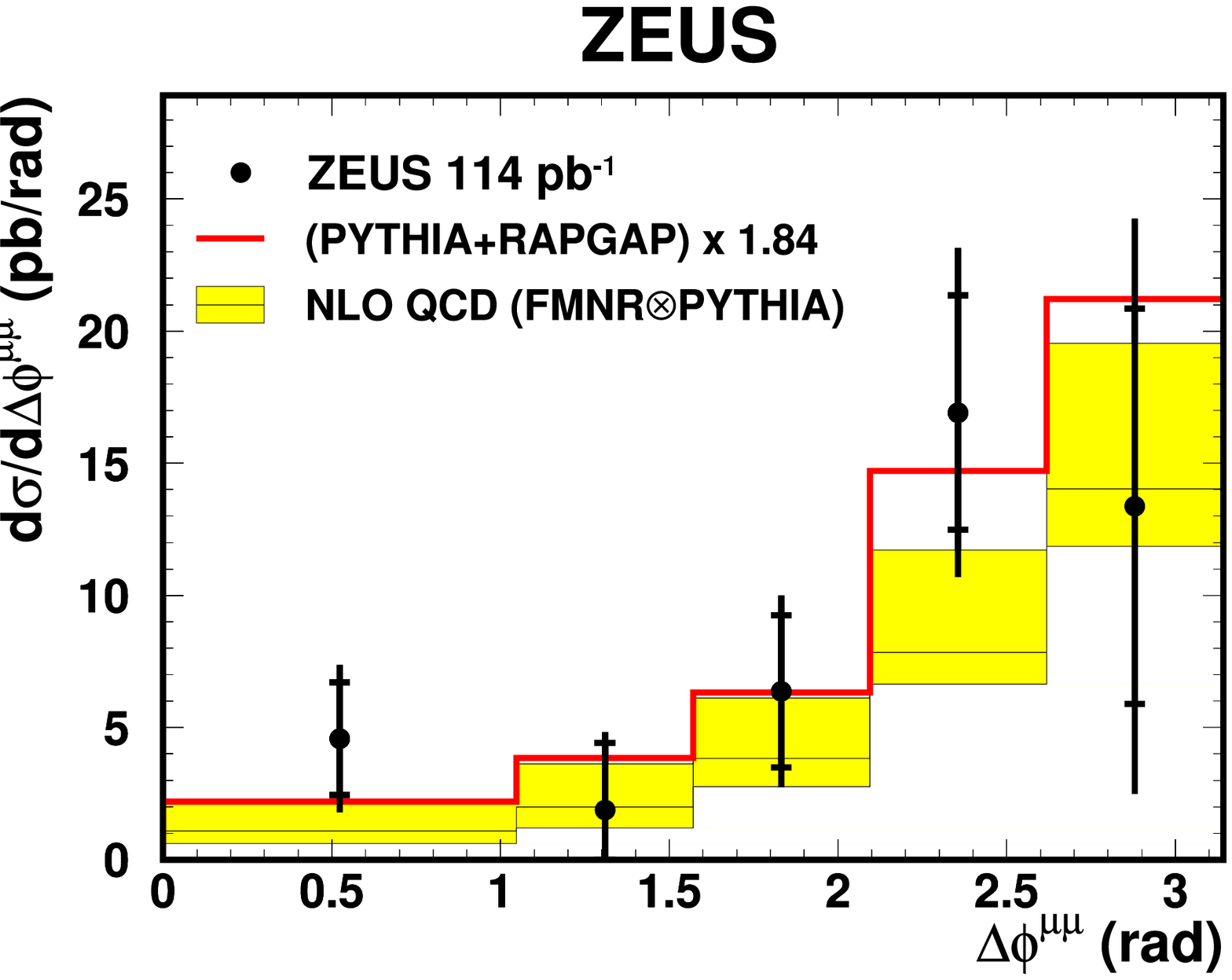}
  }
  \vskip 1cm
  \end{center}
  \caption{
    Cross-section $d\sigma/d\Delta\phi^{\mu\mu}$ for $b\bar b$ events in 
    which the muons originate from different 
    $b$ quarks, with $p_T^\mu>1.5$ GeV and $-2.2<\eta^\mu<2.5$ 
    for both muons.
    The data (solid dots) are compared
    to the scaled sum of the predictions by the LO+PS generators 
    {\sc Pythia} and {\sc Rapgap} (histogram)
    and to the NLO QCD predictions from 
    {\rm FMNR}$\otimes${\sc Pythia} (band).
  } 
  \label{fig7}
\end{figure}

\begin{figure}[htbp]
  \begin{center}
    \begin{picture}(160,70)
      \put(0,-3){\epsfig{file=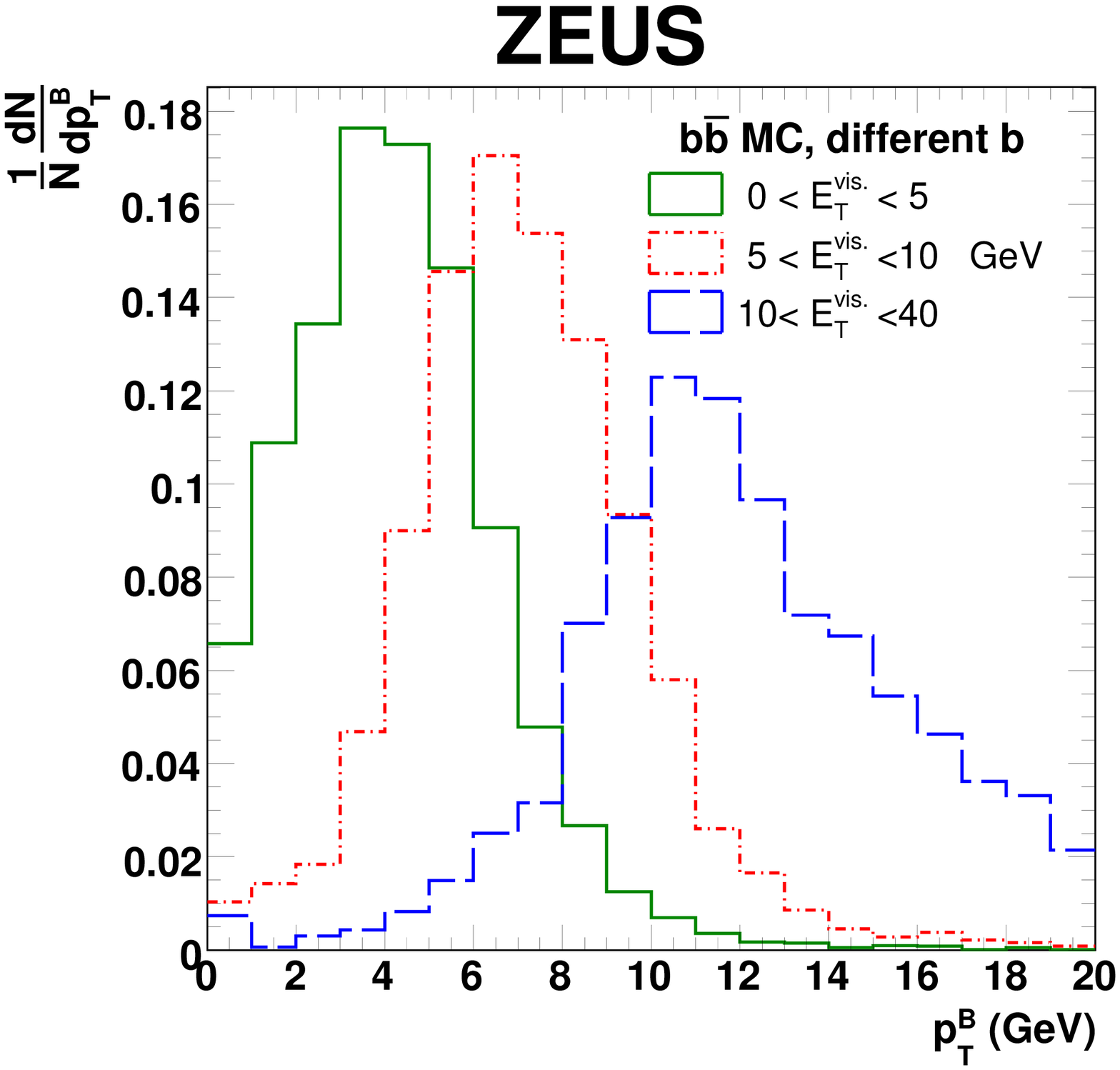, width=7.8cm, clip}}
      \put(78,-3){\epsfig{file=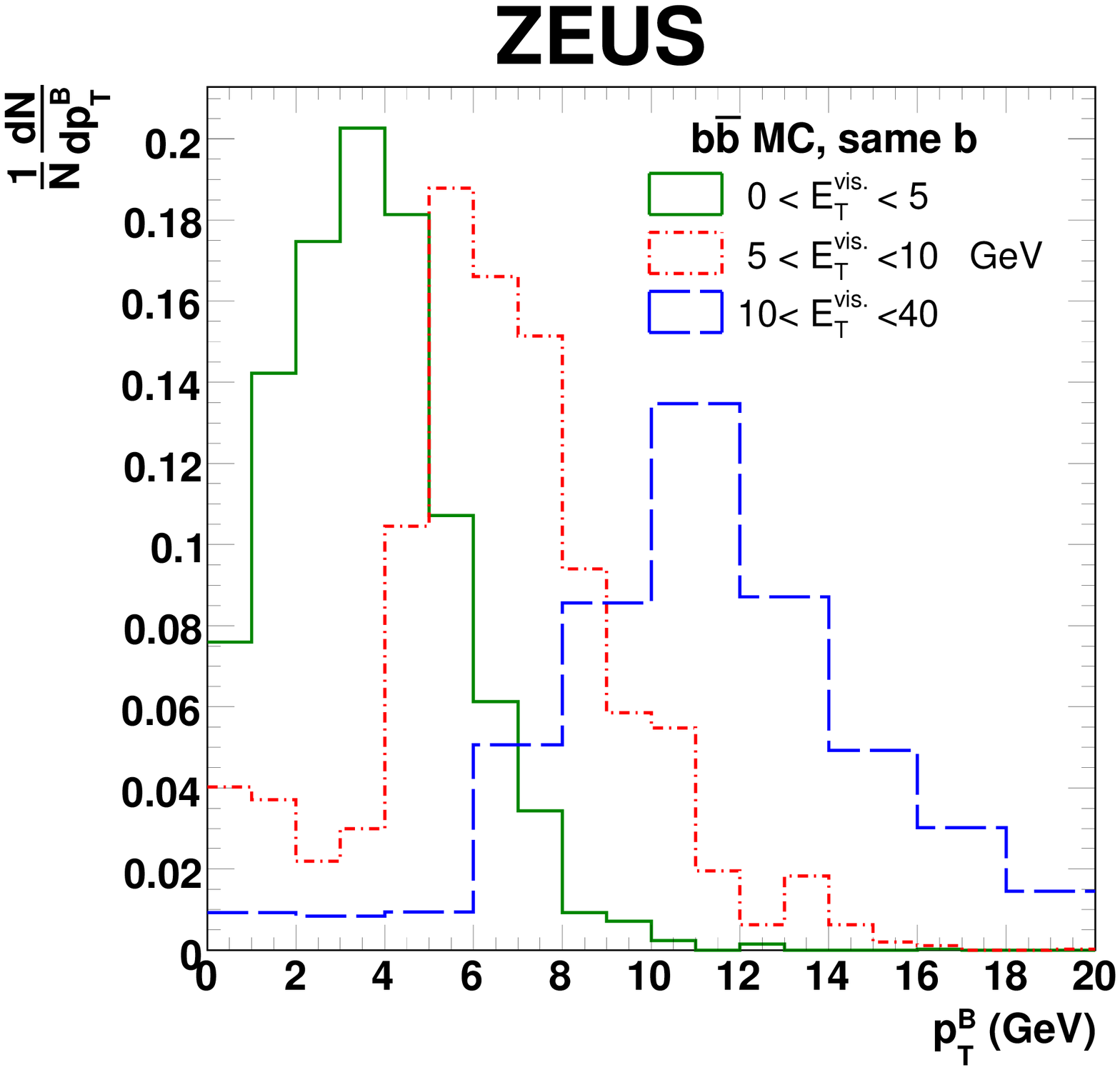, width=7.8cm, clip}}
      \put(18,57){(a)} 
      \put(96,57){(b)} 
    \end{picture}
  \end{center}
  \caption{
    Distribution of the true $p_T$ of the parent $B$ hadron for muons from (a)
    different $b$ quarks or from (b) the same $b$ quark, for the three  
    $E_T^{vis}$ bins indicated in the figures.
    For the definition of $E_T^{vis}$, see Section \ref{sec-hadron}.
    } 
  \label{fig:ptBbins}
\end{figure}

\begin{figure}[htbp]
  \begin{center}
    \begin{picture}(160,160)
      \put(37,89){\epsfig{file=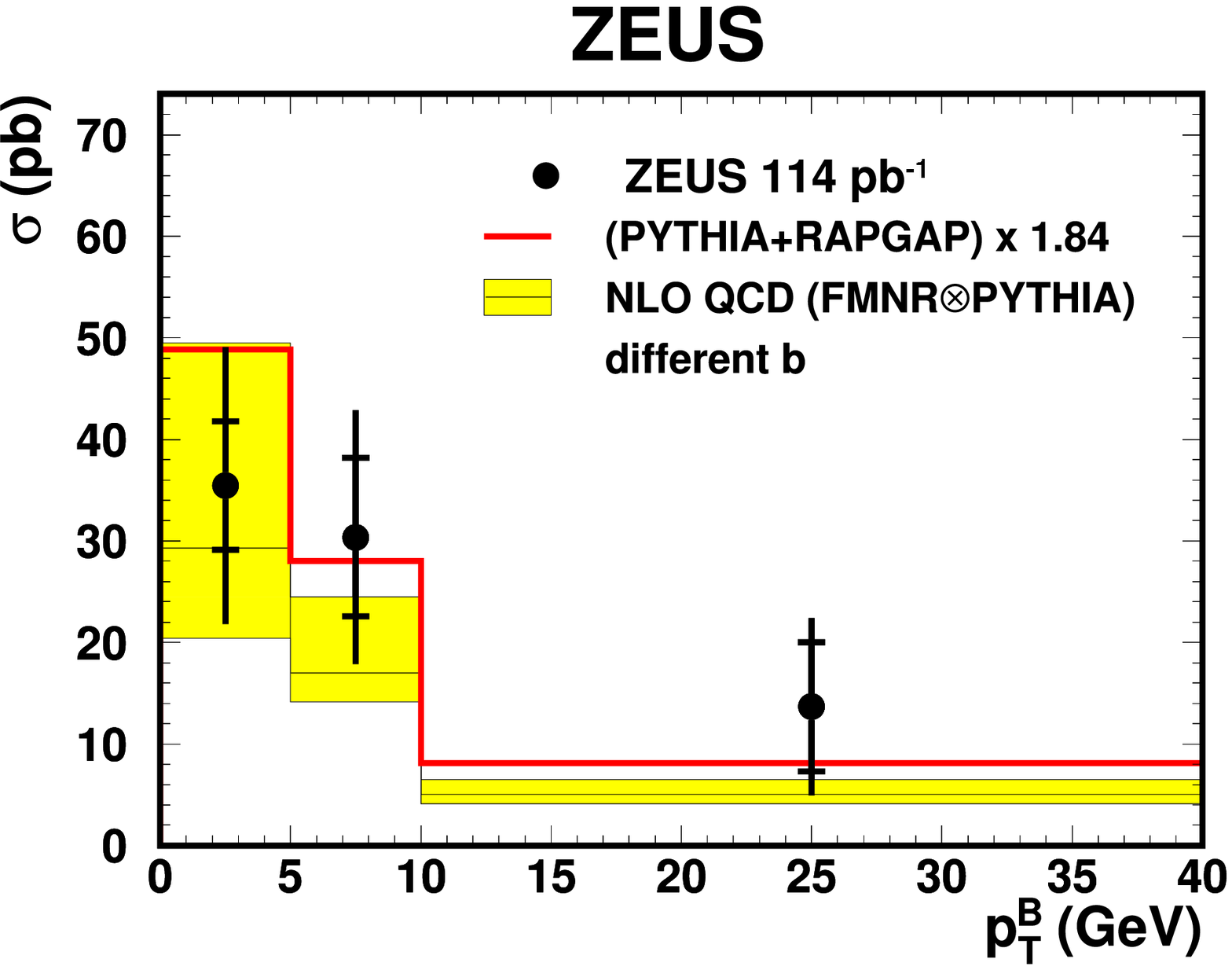, width=8.3cm}}
      \put(37,9){\epsfig{file=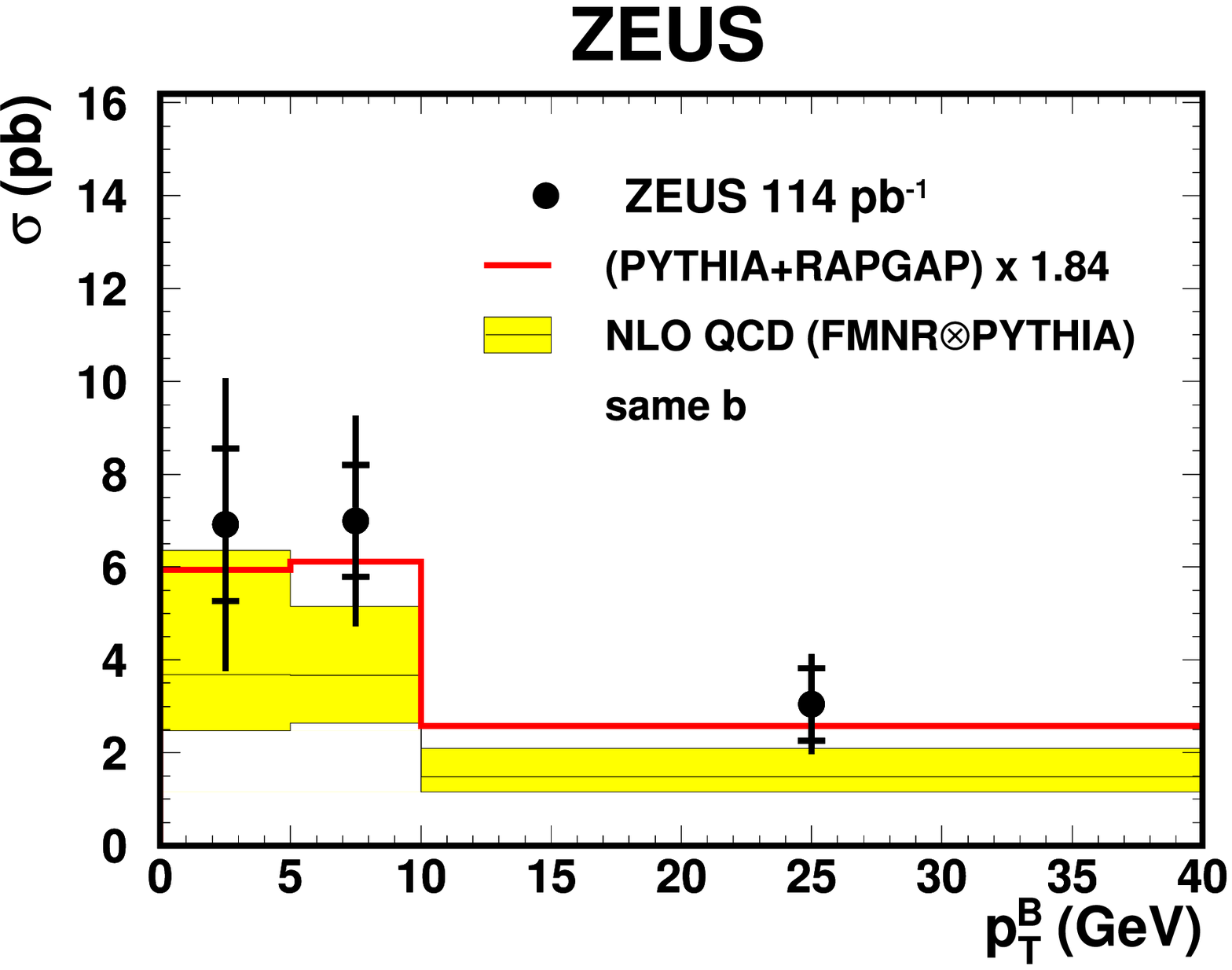, width=8.3cm}}
      \put(43,140){(a)} 
      \put(43,60){(b)} 
    \end{picture}
  \end{center}
  \caption{
    Visible cross section for parent B hadrons from events containing two 
    muons satisfying the cuts for the total cross-section measurement,
    and in which both muons originate from a different (a) or from the 
    same $b (\bar b)$ quark (b), in three bins of $p_T^B$. 
    There are two entries per event for (a), and one entry per event for (b).  
    The data (solid dots) are 
    compared to the scaled sum of the predictions by the LO+PS generators 
    {\sc Pythia} and {\sc Rapgap} (histogram) and to the NLO QCD predictions 
    from {\rm FMNR}$\otimes${\sc Pythia} (band).
  } 
  \label{fig:ptB}
\end{figure}

\begin{figure}
\begin{center}
\includegraphics[totalheight=12cm]{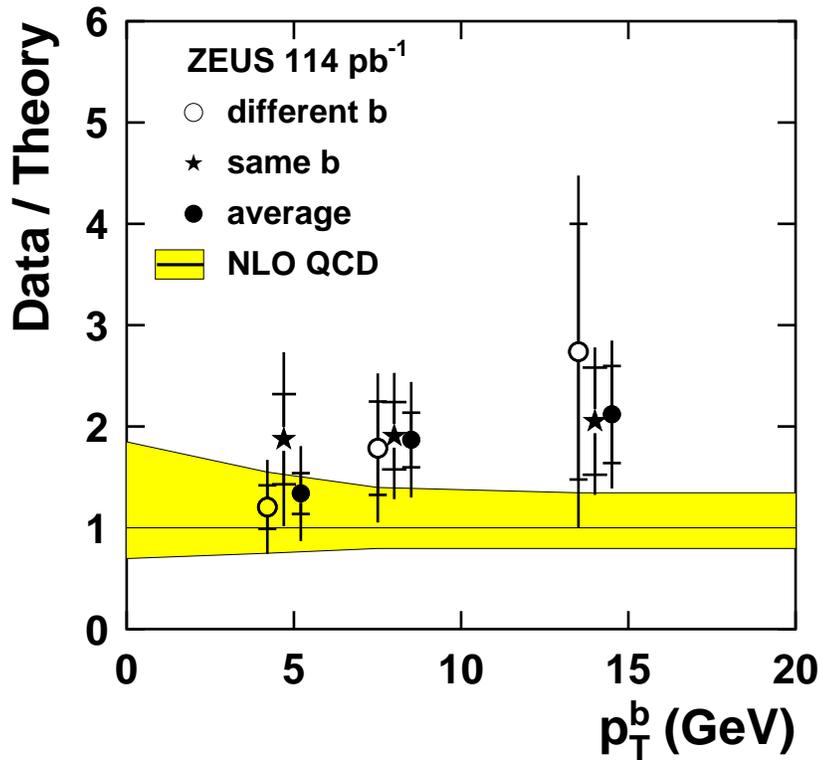}
\end{center}
\caption{
Data/NLO ratio for cross sections from different
$b$ quarks (open circles) compared to measurements from the same $b$ quark 
(stars) and their average (filled circles). The value for each $E_T^{vis}$ 
(or $p_T^B$) interval (0-5,5-10,10-40 GeV) is quoted at the median $p_T$ of 
the parent $b$ quarks in events satisfying all detector level cuts 
(4.7,8.0,14.0 GeV). The three points for each $p_T^b$ value 
are shown slightly shifted in $p_T^b$ for clarity.}
\label{fig:ptbbins}
\end{figure}

\begin{figure}
\hspace{-1cm}
\includegraphics[totalheight=12cm]{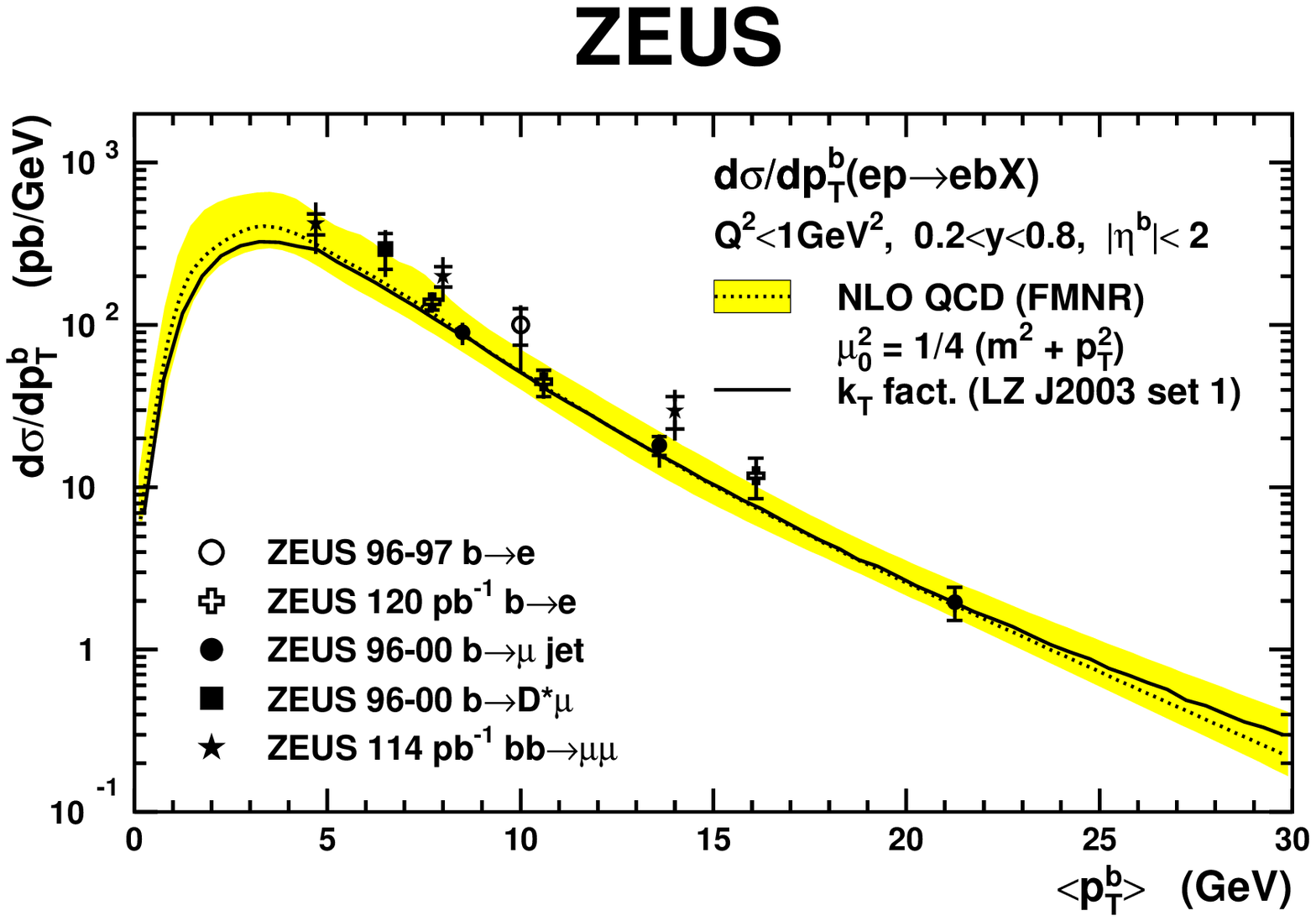}
\caption{ 
Differential cross section $d\sigma/dp_T^b$ of this analysis
(stars) compared to previous ZEUS measurements
(other symbols), FMNR NLO QCD predictions (band), 
and predictions from the $k_T$ 
factorisation approach (thick line).}
\label{fig9}
\end{figure}

%
%
\end{document}